\documentclass[%
 reprint,
nofootinbib,
 amsmath,amssymb,amsfonts,
 aps,
prd,
]{revtex4-2}

\usepackage{graphicx}
\usepackage{dcolumn}
\usepackage{bm}
\usepackage[normalem]{ulem}
\usepackage{xcolor}
\definecolor{brickred}{rgb}{0.8, 0.25, 0.33}

\usepackage[flushleft]{threeparttable}
\usepackage{longtable}

\begin{document}

\preprint{APS/123-QED}

\title{Adaptive friends-of-friends algorithm for identifying gravitationally bound cosmological structures}

\author{Prateek Gupta$^{1,2}$}
\email{prateekgidolia@gmail.com}
\author{Surajit Paul$^{3,1,4}$}
 \affiliation{$^{1}$Department of Physics, Savitribai Phule Pune University, Pune-411007, INDIA}
 \altaffiliation{$^{2}$ Indian Institute of Astrophysics, Bengaluru, INDIA}
 \affiliation{$^{3}$Manipal Centre for Natural Sciences, Centre of Excellence, Manipal Academy of Higher Education, Manipal, Karnataka 576104, India}%
\altaffiliation{$^{4}$Raman Research Institute, Bengaluru, INDIA}

\begin{abstract}
The Universe at the present epoch is found to be a network of matter over-dense and under-dense regions. Usually, the over-dense regions are dominated by the dark-matter (DM) filaments where massive gravitationally bound structures such as groups and clusters of galaxies form at the nodes. At the cosmological time scales, the baryonic matter follows the flow of DM only and together form the cosmic web. To date, this picture of the Universe is best revealed through cosmological large-volume simulations and large-scale galaxy redshift surveys, in which, the most important step is the appropriate identification of structures. So far, these structures are identified using various group finding codes, mostly based on the friend of friends (FoF) or spherical over-density (SO) algorithms. Although, the main purpose is to identify gravitationally bound structures, surprisingly, the mass information has hardly been used effectively by these codes. Moreover, while it is an established fact that the bound structures can best be formed at some particular mass over-dense regions and practically the large-scale structures are barely spherical in shape, the methods used so far either constrain the over-density or use the real unstructured geometry only. Even though these are key factors in the accurate determination of structures-mass information that can precisely constrain the cosmological models of the Universe, hardly any attempt has been made as yet to consider these important parameters together while formulating the grouping algorithms. In this paper, we present our proposed algorithm called the Measure of Increased Tie with gRavity Order (MITRO) which takes care of all the above-mentioned relevant features and ensures the bound structures by means of physical quantities, mainly mass and the total energy information. Unlike the usual FoF method where a statistically chosen single linking length is used for all grouping elements, we introduced a novel concept of physically relevant arm-length for each element depending on their individual gravity leading to a distinct linking length for each unique pair of elements. This proposed algorithm is thus fundamentally new that, not only able to catch the gravitationally bound, real unstructured geometry very well, it does identify it roughly within a predefined physically motivated density threshold. Such a thing could not be simultaneously achieved before by any of the usual FoF or SO-based methods. We also demonstrate the unique ability of the code in the appropriate identification of structures, both from large volume cosmological simulations as well as from galaxy redshift surveys, highlighting the fact that it mitigates a few shortcomings of the basic FoF and SO algorithms and strengthens the foundation of clustering or halo-finding methods in general.
\end{abstract}

\maketitle
\section{\label{intro}Introduction}
The abundance of large-volume cosmological simulations (e.g., Millennium \cite{Springel2005}, Millennium-II \cite{Boylan-Kolchin2009}, Bolshoi \cite{Klypin2011}, Cosmo-OWLS \cite{LeBrun2014}, IllustrisTNG-300 \citep{Springel2018} etc.,) as well as the accessibility of increasingly large observational surveys (e.g., SDSS \citep{Stoughton2002, Ahumada2020}, KiDS-1000 \citep{Kuijken2019}, Euclid \citep{Laureijs2011, Euclid2020}, LSST \citep{Ivezic2019, LSST2012} etc.,) in the recent times have enabled the extensive study of dynamics and evolution of structures at large scales ($\gtrsim 100$~Mpc) of the Universe. Concurrently, they provide a unique opportunity for quantitative tests for the structure formation theories along with unravelling the fabric of the Universe. The key structural elements that truly reveal the evolutionary scenario of the Universe are the dark matter (DM) halos in cosmological simulations or the clusters of galaxies in large galaxy surveys. With the appropriate structure-mass information, such as the halo mass function \cite{Press1974, Jenkins2001, Tinker2008}, one can precisely constrain the standard cosmological model parameters \cite{Bohringer2003, Allen2011, Tinker2012} or even may come up with an alternative evolutionary model for the Universe, \citep[e.g.][etc.]{Salvati2020, Klypin2021}. The cluster counts as a function of their velocity dispersion and redshift, significantly constraining the dark energy models \cite{Newman2002}. Moreover, the extraction of more accurate information about the location, mainly the position, the extent, and the bulk velocity of these structures lead to a fair understanding of their dynamics and the energy budget. Thus an obvious and very fundamental question that springs to mind is how to appositely identify these cardinal cosmic structures.

Historically, efforts have been made in this direction primarily through grouping or halo finding \cite{Press1974, Davis1985, Turner1976, Huchra1982, Zeldovich1982}. In the last few decades, many halo finders have been developed by several authors \cite[see][and the references therein]{Knebe2011, Onions2012, Knebe2013} to identify the structures and sub-structures from the copious simulated cosmological data sets. Likewise, various algorithms were developed to extract galaxy groups and clusters from the galaxy catalogue or large galaxy surveys \cite{Eke2004, Yang2005, Tago2010, Gerke2012, Pereira2017, Rodriguez2020}. Ideally, halo finders search locally over-dense, gravitationally bound systems in the matter density field (dark matter or dark+baryonic matter), either generated in the simulated realisations (N-body/hydro-dynamic or hybrid-codes) or within any large galaxy-redshift surveys. Although numerous halo finders have been developed so far, nearly all of them have their foundation either in the spherical over-density (SO) algorithm \cite{Press1974} or in the friends-of-friends (FoF) algorithm \cite{Davis1985}, otherwise, they are the derivatives or combinations of these two basic methods.

The classical SO algorithm locates the density peaks in the matter density field, the spherical shells are then stacked up around these peaks until the enclosed average density drops below a certain predefined threshold value. The threshold value is assumed either as some multiples of the mean density of the simulated volume or as the critical density of the Universe at that redshift. Consequently, this enforces a spherical geometry in SO halos. Whereas, the FoF halos are unstructured, as the FoF algorithm collects the particles that are close to each other with respect to a linking length, which is usually a fractional multiple of the mean separation of the particles in the searched volume. Certainly, each of these basic algorithms has its advantages and limitations.

While, the SO-based algorithms, by enforcing a spherical symmetry, fail to capture the real shape of the halos, undoubtedly they pick the halo centres as the density peaks, fairly well. Whereas, though the FoF algorithm is successful in discerning the real geometry of the halos to a great extent, it may not centre the halos at the density peak of the systems so accurately. This is, largely because it locates centres at the centre of mass or simply at the mean position vector of the final structure. Furthermore, the FoF algorithm is fundamentally based on a free parameter called the `linking length' ($l_{f}$), which can be fine-tuned depending on the scientific problems to be addressed. All trial values that are assumed by various authors in this scheme are as yet chosen statistically, e.g., the fraction of mean inter-particle separation in the volume of interest ($l_{f}$ = $b\bar{l}$, where $b \in (0,1)$ and $\bar{l}$ is the mean separation). The major limitation of this parameter `$b$' is, that the choice mostly has no physical relevance and may connect two nearby gravitationally unbound structures into one, if the chosen linking length is too large. Even, the widely used value, $b=0.2$, which some of the authors have claimed to roughly correspond to the halo-over-density of $\sim200$, the value often assumed for the virialized systems, was found to be not so consistent when tested using simulated data with different simulation set-up. The difference may be as large as 100-200 of the mean density as reported by \cite{Warren2006,More2011}. In general, no absolute relation has so far been established between $b$ and the over-density of the halos. This necessitates a physically motivated linking parameter that would catch the real gravitationally bound systems. Ironically, these methods ignore the mass information of individual elements of the systems, while connecting them, overlooking the fact that at large scales, gravitation is the only and strongest entity that effectively holds the elements together.

In this paper, we present our efforts to design an advanced halo-finding method comprising of a few novel features along with addressing aforesaid limitations of previously developed algorithms, invoking physically motivated grouping parameters. The paper is laid out as follows. After introducing the article in section~\ref{intro}, we begin with the development of the proposed algorithm and illustrate the methods to estimate the associated basic physical properties of the identified halos in section~\ref{method}. In the first part of section~\ref{result}, we elaborate on the application of the proposed halo-finder to the simulated data sample. In the later part of the section, we present the applicability of the algorithm (with a few essential modifications) to the observed galaxy catalogue of the Sloan Digital Sky Survey (SDSS) in the region of the SARASWATI supercluster. Finally, we summarise our findings and conclude the paper in section~\ref{summary}.

\section{\label{method}Methods}
Our primary motivation behind developing a new halo-finding algorithm is to connect it to more physical parameters pertinent to both the simulated as well as observational data. Moreover, the algorithm should have powerful features from the basic FoF and SO methods, and intrinsically it should be able to overcome some of the crucial shortcomings in them. For instance, it should avoid forcing a spherical symmetry in describing the halos, rather it should keep the unstructured geometry as the DM halos are not necessarily following the spherical symmetry during their formation through accretion or violent mergers inside the cosmic web. Furthermore, the code should follow the physically more relevant over-density parameter in describing the boundary of the halos, reducing the usual dependence on many non-physical free parameters.

The proposed halo finder namely the Measure of Increased Tie with gRavity Order (MITRO) has analogous basic features of the FoF algorithm. In the framework of FoF, the nearest neighbour of, say the $i^{th}$ element (here, DM particle or galaxy) is searched within a predefined radius known as the linking length, $l_f$. The $j^{th}$'s elements in the examined volume, that are located within the distance $l_f$ with respect to the $i^{th}$ element, are considered as its friends (the direct friends). Likewise, a further search is done for the new friend(s) of these direct friends of the $i^{th}$ element with the same linking length around them and are known to be the indirect friend(s) of the $i^{th}$-element. The process is repeated until no new friend(s), direct or indirect, are found for the $i^{th}$-element in the searched volume. The final configuration of these groups of friends (direct and indirect) will be considered as a DM halo (in the case of cosmological simulations) or a cluster of galaxies (in the case of galaxy surveys). In this setup, the FoF halos primarily depend on the choice of a free parameter, called the linking length. Choosing quite a small (or large) linking length will potentially lead to the formation of smaller (or bigger) size halos. Moreover, the enclosed over-density of these halos with such an arbitrarily chosen linking length provides no absolute or even any general empirical relationship among them (see \cite{Frenk1988, Lacey1994, Audit1998, Warren2006} and references therein).

\subsection{\label{model}MITRO algorithm}
The proposed algorithm MITRO introduces a new concept of adaptive and unique linking length between the pairs of particles in the search field. To visualise the basic concept of this novel algorithm, let us take a practical example. Imagine that a flock ($\mathbb{F}$) of people sitting randomly on the ground and we need to search the groups of people in $\mathbb{F}$ who can hold each other's hands. The simplest way to do this is to apply the basic FoF method. In which, first a connecting length is set, which in FoF is usually a weighted fraction of the mean separation between the two people in the flock $\mathbb{F}$ and then the aforementioned steps of the FoF algorithm are applied. As we have already mentioned, the resulting FoF groups are very much sensitive to the chosen linking length and therefore need a careful determination of the parameter to produce meaningful results. For example, in this case, if we choose twice the average arm length as the connecting length for everyone in the flock, many people with shorter arm lengths will be counted in the group who in actual cases will hardly be able to hold themselves together to form a connecting group. In fact, arm length is an individual's physical property that depends on various parameters such as age, height etc., therefore, the assumption of a fixed linking length would make the group non-physical.

The issue described above is not just a problem specific. It has been noticed several times that to achieve a meaningful grouping in galaxy survey data using FoF, the adopted linking length may be very large \cite{Tago2010, Tempel2014}. Since a single FoF linking length is used for the entire data, irrespective of the mass of individual galaxies in the sample, it is highly likely that many low-mass galaxies in the sample may not have gravitational influence to such a distance to make a bound system. In such a case, the groups that are chosen by the FoF algorithm hardly represent a gravitationally bound system in a physical sense.

To address this issue, one needs to rectify the friending method which is so far based on a fixed linking length. For simplicity, let us first assume that each person in the said flock has as many arms as s/he demands to connect with the neighbours. To get connected, they should stretch out their arms and try to hold the hands of their neighbours without changing their position on the ground. Here, two people are called their direct friends if they are able to hold each other's hands. And the people are called friends of friends (or indirect friends) if they are connected through intermediate friends. In this way, the resulting interconnected configuration of friendship circles is said to be the physically connected groups of people.

In an analogous situation as stated above, the MITRO algorithm addresses the same problem in a different and more efficient way. Let us assume that in a given cosmological data set, there are two particles (either the dark matter particle or the galaxy), $A$ and $B$. Naturally, they may have different masses, say $M_{A}$ and $M_{B}$, and therefore in general different arm lengths of $L_{A}$ and $L_{B}$, respectively, depending on the effective field of attraction of these gravitating bodies as elaborated in section~\ref{armlen}. Accordingly, the linking length between them will be $L_{AB} \equiv L_{A} + L_{B}$. Now, if there is a third particle, say $C$, of mass $M_{C}$ and arm length $L_{C}$, the possible linking length between $A\;\&\;C$ will be $L_{AC}$ ($\equiv L_{A} + L_{C}$) and the same between $B\;\&\;C$ will be $L_{BC}$ ($\equiv L_{B} + L_{C}$). So, the significant characteristic of the MITRO algorithm is the introduction of the concept of arm-length and the existence of as many linking lengths as the number of unique combinations of pairs of elements in the volume of interest. Overall, one may consider MITRO as a cousin of FoF with the major difference that the former adopts the concept of distinct linking length for each unique pair of particles instead of a common linking length for all the pairs, as is the basis of the latter.

\begin{figure}
	\includegraphics[width=1.0\columnwidth]{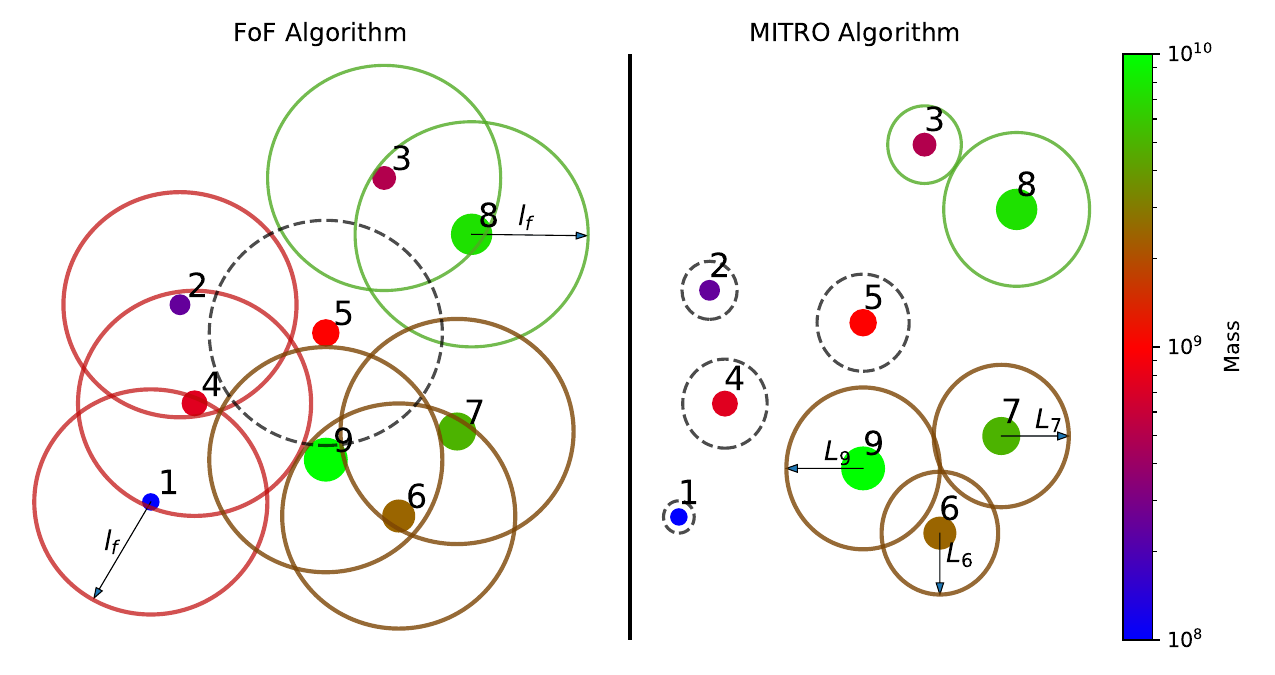}
    \caption{The colour dots in the figure represent the particles in 2-dimensional Euclidean space, where the colour code and the size of the colour dots are indicative of the mass of the respective particle. \textit{\textbf{Left panel}}: demonstrates the clustering with FoF algorithm. Big circles around the colour dots represent the linking length ($l_f$) of each particle. Whereas, \textit{\textbf{Right panel}}: demonstrates the clustering with MITRO algorithm. Here, big circles around the colour dots represent the arm-length ($L_i$) of the individual particle based on its mass property. In both panels, the dashed circles represent the isolated particles while the solid circles are representing those who found their friend(s) or companion. The different colours for the solid circles are to differentiate the identified groups. \textit{FoF groups}: $\{$P1, P2, P4$\}$, $\{$P6, P7, P9$\}$, and $\{$P3, P8$\}$. \textit{MITRO identified groups}: $\{$P6, P7, P9$\}$, and $\{$P3,  P8$\}$}
    \label{fig:algodiff}
\end{figure}

Now, while searching for friends, any two elements are said to be connected or are direct friends, if and only if the Euclidean distance ($D_{ij}$) between them is less than or equal to the sum of their individual arm-lengths, i.e., $D_{ij} \leqslant L_{i} + L_{j}$. In this way, one computes the direct friend(s) of $i^{th}$-element. Subsequently, the friend(s) and further friend(s) of their friend(s) of $i^{th}$ particle are altogether collected in a set. Finally, each disjoint set of friends will represent a group in the search volume. Moreover, this method also keeps in mind that the friendship between two individuals is not just the efforts of one, instead, the effort should be from both sides. A pictorial illustration of the above-discussed method can be seen in the right panel of Fig.~\ref{fig:algodiff}.

In Figure~\ref{fig:algodiff}, we pictorially mimicked a specific clustering scenario with FoF (left panel) and with our proposed algorithm (right panel). Nine elements with different masses are randomly placed in a two-dimensional Euclidean space. For better representation, the particles of different masses are indicated by the filled dots of different sizes with dots getting larger for the higher masses. Colours are used for making them distinctly visible. In the left panel, we applied the FoF algorithm with a fixed linking length ($l_f$), computed as the mean separation between the particles in the represented area. As per the usual FoF algorithm, as stated earlier, we obtained three different groups: $G^{F}_{1}:\{$P1, P2, P4$\}$, $G^{F}_{2}:\{$P3, P8$\}$, and $G^{F}_{3}:\{$P6, P7, P9$\}$, and an isolated particle, P5, surrounded by the dashed line circle. However, for the same set of particles, when we applied our proposed MITRO algorithm, we obtained only two different groups: $G^{M}_{1}:$ $\{$P3, P8$\}$, and $G^{M}_{2}:$ $\{$P6, P7, P9$\}$ and rest as the isolated particles, as demonstrated in the right panel of Fig.~\ref{fig:algodiff}. Here, the arm-lengths ($L_i$) have been computed by scaling with the mass of individual particles, as proposed in the MITRO algorithm. Naturally, elements differing in mass have different arm-lengths. Since, by definition of MITRO, an examined pair becomes friends only when the arm of one overlaps or at least touches the other and thus accordingly the groups are formed. The results obtained in MITRO seem physically more relevant. Though FoF flags particles $\{$P1, P2, P4$\}$ as groups, namely $G^{F}_{1}$, these are very small masses compared to the other elements in the set and in the actual case may not have enough potential to make a bound system and therefore our proposed algorithm has truly flagged them out as isolated particles. We will further elaborate on this issue in various sections throughout the paper.

\subsubsection{\label{armlen}The concept of arm length}
Since the major goal of running halo finder codes in large-scale matter distribution is to find the gravitationally bound systems, the first thing to ensure is that the elements of a group are under the gravitational field of influence. The basic theory of gravity says that particles with different masses will have different fields of influence. Therefore, in a system of particles with different masses, the heavier mass particles have a greater impact on the trajectories of other nearby smaller mass particles. This seeds the idea that in a gravitating system, the most effective physical parameter, which is definitely the mass of the elements (DM particles or galaxies), should determine the distance of influence. We call this as the effective radius or the arm-length of the mass element. This in turn will decide the final possible linking length between any unique pair of particles in our proposed algorithm. Linking length is thus no more a free parameter, unlike in FoF, but rather an intrinsic physical property of the individual pair of particles in the system.

As a preliminary test step, we considered the gravitational potential as the measure of influence and determined the arm length. For any arbitrary mass $M$, the associated gravitational potential ($V$) at any distance $r$ is given by $V = -\frac{GM}{r}$, where $G$ is the Universal gravitational constant. The maximum arm length can therefore be taken as the distance from the particle where its influence almost vanishes, which is, by definition, at an infinite distance. However, a finite arm length is essential to get a meaningful clustering. For simplicity of calculation, one can attempt to get a distance at which the increment in the gravitational potential, i.e., $\frac{\partial V}{\partial r} \equiv \frac{GM}{r^{2}}$, becomes considerably low. Assuming the radial rate of change of gravitational potential at a distance $R$ (as $F_{R}$) to be such a limit, we get
\begin{eqnarray}
F_{R} = \left\vert \frac{\partial V}{\partial r} \right\vert_{r=R}  \equiv \frac{GM}{R^{2}}
	\label{eq:eq1}
\end{eqnarray}
This leads to an arm-length of
\begin{eqnarray}
    R = \sqrt{\frac{GM}{F_{R}}} \approx 3.73\times10^{-9} \; \text{kpc} \left( \frac{M}{\text{M}}_{\odot} \right)^{1/2} F^{-1/2}_{R}
	\label{eq:eq3}
\end{eqnarray}
For the assumed cosmological set-up and simulated data set used in this study, we found that $F_{R}\leqslant 10^{-7}$ produces a meaningful clustering\footnote{$F_{R}$ in Eq.~\ref{eq:eq3} does not express any direct physical quantity, rather it signifies the slope of the gravitational potential at some arbitrary distance R due to the particle of mass M. In other sense, $F_{R}$ may be realised as force (dynes, in c.g.s. units) applied on a unit mass of test particle placed at distance R from mass M.}. However, the number (i.e., $\leqslant 10^{-7}$) obtained here is through trial and error and needs some robust method to compute such a limit, which is out of the scope of this study. Moreover, it should also be physically motivated. This led us to define the arm-length in a generic way by relating it to the basic cosmological definition for the spherically collapsed objects.

We define the arm-length as the radius, $R_{\Delta}$, of a sphere around the particle of mass $M$ within which we uniformly distribute the mass such that the mean density of the sphere becomes $\Delta$ times the critical density (${\rho_c}_{z}$).

\begin{eqnarray}
    \frac{M}{\frac{4}{3} \pi R^{3}_{\Delta}} =  \Delta \times {\rho_{c}}_{z}
	\label{eq:eq4}
\end{eqnarray}
Here, the critical density, ${\rho_c}_{z}$ corresponds to the redshift of the particle and the assumed cosmology. The arm-length is sensitive to the choice of $\Delta$, however, its value can be limited to any physically motivated situation. For instance, if one needs to search for the structures at large scales ($\gtrsim$ a few $10$'s of Mpc), $\Delta \in [2,2000]$. The arm-length for each particle can be determined as,
\begin{eqnarray}
    R_{\Delta} = \left[ \frac{M}{\frac{4}{3} \pi \Delta {\rho_{c}}_{z}}  \right]^{1/3}
	\label{eq:eq5}
\end{eqnarray}
An important advantage of over-density mediated arm-length is that it also enables us to fairly constrain the enclosed over-density ratio ($\overline{\rho}/\rho_{c}$) of the identified halos using the MITRO halo finder (discussed later in sec.~\ref{Compactness}).

\subsubsection{\label{sec:unbinding}Unbinding process}
Locating halos using a spatial clustering algorithm usually ensures the positional proximity of the halo elements. However, such clustering does not guarantee that all the elements in the halos are energetically bound to the system. Rather, the position-space-based halo finder will always include a few high-velocity particles which could be dynamically unrelated to the system \cite{Knebe2013}. Identification and removal of such unbound particles in a halo are known as the unbinding procedure. Implementation of the unbinding procedure allows us to make the halo catalogues free from significant contamination of spurious small objects and therefore an essential practice for filtering sub-halos out of the main halo as well as computing more accurate halo centres and its bulk velocities \cite{Onions2013}. In our approach to this process, a particle is said to be unbound to the system (i.e., a halo identified by MITRO), if its rest-frame velocity is greater than or equal to the escape velocity computed at the particle position. So, in the formed cluster $\mathbb{C}$, with particles, $P_{j},\; j \in \mathbb{C}$; the particle $P_{j}$ is flagged as an unbound particle if the kinetic energy of $P_{j}$ is greater than or equals to the total gravitational potential of the system at the particle position due to the rest of the particles in $\mathbb{C}$. Such particles will then be de-flagged from the member list of $\mathbb{C}$. Since in the large-scale structures of the Universe, material flows along the filaments connecting the galaxy clusters and the whole network are in general in motion, to compute the kinetic energy of the particles, we have first corrected the velocity components ($v_{i}$'s) by subtracting the bulk flow of the cluster. The kinetic energy of `$j^{th}$' particle in the rest frame of cluster ($\mathbb{C}$) is therefore given by;
\begin{eqnarray}
    KE_{j} = \frac{1}{2} M_{j} \sum_{i \in \{x,y,z\}} \left( v_{ij} - V^{bulk}_{i} \right)^2 \;\;\;\;\; \forall \;\;\;\; j \in \mathbb{C}
\end{eqnarray}
where $M_{j}$ is the mass of $j^{th}$ particle in $\mathbb{C}$, and $V^{bulk}_{i}$ denotes the $i^{th}$ velocity component of the bulk flow of the cluster $\mathbb{C}$ which is evaluated as;
\begin{eqnarray}
    V^{bulk}_{i} = \sum_{j \in \;\mathbb{C}} v_{ij} M_{j} \;\;\;\;\; \forall \;\;\;\;\;\; i \in \{x,y,z\}
\end{eqnarray}
Although for a faster computation of gravitational potential one may use the tree code, particle number being not so large ($\leqslant 10^{6}$) in our specific problem, as well as, for a more precise evaluation of the same, we compute it as;
\begin{eqnarray}
    PE_{j} = \sum_{k \neq j} G \frac{M_{j} M_{k}}{|\vec{r}_{jk}|} \;\;\;\;\; \forall \;\;\;\; k \in \mathbb{C}
\end{eqnarray}
Now, for each particle $P_{j}$ in cluster $\mathbb{C}$, wherever $KE_{j} \geqslant PE_{j}$, it will be untagged from the membership of $\mathbb{C}$. We iterate the process of unbinding till all such unbound particles are removed from the system $\mathbb{C}$.

Nevertheless, switching on the unbinding process is not an absolutely necessary step for a halo-finder. It mostly depends on the scientific problem that one deals with. The halo properties that are based on gravitating matters only, such as gravitational lensing, X-ray properties, Sunyaev-Zeldovich etc, actually require all the particles that are positioned in the configuration, even if they are not bound to the system \cite{Onions2013}. Moreover, if one is interested in studying the diffuse streaming or turbulence properties inside the group medium, the running unbinding process may even turn out to be detrimental as such calculations rather than have an absolute requirement for those high-velocity particles that are present in the system. Consequently, we introduced unbinding process as an optional step in our code.

\subsection{\label{basic_halo_properties}Computed halo properties in MITRO-algorithm}
\subsubsection{\label{positioning}Halo position}
The successful determination of precise representative halo centres is an important task of a halo finder. This is mainly because of the fact that almost all basic halo properties such as radius, mass, bulk velocity etc., are primarily computed on the basis of the chosen halo centres. In our proposed algorithm, we do it by implementing two different concepts; (i) the commonly used mass-weighted average position of all member particles of a halo i.e., $r_{com}$ (the centre of mass, CoM position, hereafter), and (ii) the position of the halo member at the lowest gravitational potential i.e., $r_{lgp}$ (the lowest gravity point, LGP position, hereafter), as the halo centre. The latter one will be our main method to compute the halo positions, since this method was found to provide the halo centres closer to the actual matter density peak in the halo, in comparison with the former one as determined by the centre off-set study (see more detail in section~\ref{Position_offset}).

\subsubsection{\label{halo_over_density}Halo over-density ratio}
In large-scale structures, the over-density ratio is a key parameter that helps in defining roughly the regions that are successfully able to withstand the Hubble expansion and may eventually be collapsed to bound objects when their over-density reaches a certain critical value \cite{Press1974}. In our proposed algorithm, we compute the over-density ratio of a halo as the ratio of the matter density in the volume enclosed by the halo to the critical density of the Universe at the halo redshift. In our method, no spherical geometry for the halos has been enforced, rather it keeps the intrinsic shape of the surface that encloses all the halo-fellow particles.

We construct the envelope which encloses halo members using the Convex Hull algorithm. For any given set of points $\mathcal{X}$ in 3D-Euclidean space, the Convex Hull algorithm selects the smallest number of convex set of points ($Q_{hull} \subset \mathcal{X}$) which encloses the given set \textbf{$\mathcal{X}$} in a closed convex surface, the minimum possible volume enclosing $\mathcal{X}$. For our case, we make use of the SciPy Quick-hull Algorithm \cite{Barber1996} to construct the Convex Hull and subsequently compute the corresponding volume, \textbf{$\mathcal{V}$}, within the convex surface planes. However, in the MITRO algorithm, the set of points \textbf{$\mathcal{X}$} is not simply the set of positions of the halo members as used in the usual Convex Hull set-up. In the proposed algorithm, we assign arm-length to the individual particles and therefore they assume a spherical shape of radius equal to their respective arm lengths (sec.~\ref{armlen}). Since these spheres will have non-negligible sizes, to meticulously compute the volume \textbf{$\mathcal{V}$}, one needs to apply the Convex Hull for a set of spheres instead of a set of points. Furthermore, these spheres being of different sizes may even overlap with each other, demanding a very complex and modified Convex Hull method. To avoid such complexity, as it is out of the scope of this work, we have rather used a heuristic approach. We made a new set of points \textbf{$\mathcal{X^\prime}$} taking the point on the spherical surface of each halo member that is farthest from the centre (LGP centre) of the host-halo. In this way, instead of applying Convex Hull on the set of halo member positions, we apply it on this new set of farthest points i.e. \textbf{$\mathcal{X^\prime}$} and compute the volume \textbf{$\mathcal{V}$} of the halo.

With this, the mean density ($\overline{\rho}$) of the halos is then computed as the ratio of the total enclosed mass ($\Sigma M_{i}$) to its volume, \textbf{$\mathcal{V}$}. The over-density ratio ($\wp_{halo}$) is therefore defined as;
\begin{eqnarray}
    \wp_{halo} = \frac{\overline{\rho}}{\rho_{c}}\;\;\;\; where\;\;\;\;
    \overline{\rho} = \frac{1}{\mathcal{V}} \sum_{i \in \; \mathcal{X}} M_{i}
\end{eqnarray}

\subsubsection{Halo radius and mass}
The halo radius in our code is calculated in a very straightforward way. We report the distance between the LGP centre and the farthest member of the halo as the radius or maximum radius, $R_{max}$, more precisely, the farthest point in the Convex Hull computed enclosing surface with respect to the LGP centre. However, in highly mass-resolved simulations, the mass of DM particles being small, the arm-length is negligible compared to $R_{max}$ and therefore searching the positions of the farthest members will be enough in such cases. Contrary to this, the presence of massive galaxies as the cluster member in galaxy surveys makes the arm length non-negligible, and thus in such cases, the farthest Convex Hull surface would be considered for measuring $R_{max}$. We further compute the mass of MITRO-identified halos by taking the sum of masses of each individual halo member element and denoting them as $M_{tot}$.

\subsubsection{\label{Compactness}Compactness parameter}
Introduction of the concept of arm-length allowed us to understand the stiffness of the configured halos by defining a new parameter called `compactness', the ratio of the measured halo over-density using Convex-Hull method, $\wp_{halo}$, (see section~\ref{halo_over_density}) to the reference over-density ($\wp_{ref} = \Delta_{ref}/2$) i.e., $C_p=\wp_{halo}/\wp_{ref}$. The two factors in the denominator of $\Delta_{ref}/2$ will be discussed in the further text. This `$C_p$' parameter figures out how closely or densely the member elements are packed in a halo compared to the search field element density. While defining the arm-length, given in Eq.~\ref{eq:eq5}, the size of the sphere around the particle position has been computed assuming that the mass of the particle is uniformly distributed in the sphere of radius $R_{\Delta}$, such that the enclosed density is $\Delta$ times the critical density ($\rho_{c_{z}}$). For simplicity, let us assume a gravitationally bound system with $N$ number of elements of equal masses. Equal mass implies all spheres will have the same radii. If we consider them as solid spheres, while forming a gravitationally bound structure, they may simply touch each other if placed closely, similar to the atoms in a closely packed lattice structure. Much like the lattice, we may therefore define a term packing fraction of the halo, 
\begin{eqnarray}
    P_{f} = \frac{N \times \text{Volume of each sphere}}{\text{Total volume enclosed by the structure}}
\end{eqnarray}
With a simplistic analogy of a cubic lattice filled with equal-sized spherical atoms, one gets a packing efficiency of $\pi/6 \approx 0.5$. Even if we consider spherical elements of arbitrary sizes or if we allow overlapping of spheres, the packing efficiency would only increase. This sets the lower limit to the enclosed over-density of any MITRO-identified halo to roughly $\Delta/2$. Therefore, if the DM halos are searched with $\Delta \equiv \Delta_{ref}$, the enclosed over-density of the identified halos must be $\geqq \Delta_{ref}/2$. We may therefore take $\Delta_{ref}/2$ as the density threshold in which the halos are searched and accordingly, we should set the $\Delta_{ref}$ as the specific problem may demand.

\section{\label{result}Applying MITRO-algorithm on different cosmological data sets}
The proposed MITRO algorithm is so designed that in its basic framework, it universally works on the data sets from N-body/particle simulations as well as galaxy-redshift surveys. In further sections, we tested our algorithm on cosmological simulation data by analysing the halo properties defined in section~\ref{basic_halo_properties}. Subsequently, we implemented our algorithm on observational survey data to identify large-scale structures in the Universe, specifically the data from a previously known supercluster region.

\subsection{Analysis of simulation data using MITRO algorithm}
MITRO algorithm works on any cosmological N-body simulation that has recorded at least a few basic parameters. The minimum required parameters are the particle's (1) ID, (2) mass, (3) position (in Cartesian coordinates, in its current framework) and (4) Cartesian velocity components. The particle's ID naturally works for identifying individual particles, the particle's position and mass play a significant role in the halo finding process, as discussed in sec.~\ref{model}, whereas, particle velocity is the key information for verifying the particle's association with the identified halo (see sec.~\ref{sec:unbinding}) by examining its gravitational boundedness to the system.

The simulated cosmological data sets used in this work are taken from \citet{Paul2017}. Simulations were performed using the ENZO~2.1 code \cite{Bryan2014}. ENZO is a hybrid (N-body + Hydro-dynamical), Eulerian adaptive mesh refinement (AMR) code dealing with the dark matter as particle only and baryonic matter as the fluid that realises the evolution of hydro-dynamical parameters, i.e, $\rho, P, T $, in each grid or cell of the simulated volume.

\textbf{Cosmological model:} To create the simulated realisations, a flat $\Lambda-CDM$ cosmology was assumed with $(\Omega_{\Lambda},\; \Omega_{m},\; \Omega_{b})$ = $(0.7257,\; 0.2743,\; 0.0458)$; and $h$ = $H_{0}/(100\; \rm{km}\;\rm{s}^{-1} \rm{Mpc}^{-1})$ = 0.702; a spectral index for the primordial spectrum of initial matter fluctuations, $ n_{s} =$ 1.0; and the rms amplitude of linear fluctuations in the spheres of radius $8$ Mpc $h^{-1}$, $\sigma_{8}$ = 0.812 \cite{Komatsu2009}.

\textbf{Simulation set up:} In these simulations, a part of the Universe has been mimicked within the co-moving cosmological cubic volume, V, of sides, $L_{box}$ = $128$~Mpc $h^{-1}$, filled with $64^3$ number of dark matter particles and $64^3$ cells in the parent grid. Furthermore, two static/nested child grids ($64^3$ cells) of volume V/2$^3$ and V/2$^6$ respectively, were inserted such that the centre of both the child grids coincided with the centre of the parent grid.  With the insertion of two child grids, the dark matter particle mass resolution of $7.9 \times 10^{9}$ M$_{\odot} h^{-1}$ has been achieved in the central most region of the co-moving volume ($32$~Mpc\; $h^{-1}$)$^3$. Another 4 levels of AMR on matter density parameter have been initialised in the second child grid, of volume V/2$^6$, to obtain a peak (co-moving) grid resolution of $31.25$~kpc\; $h^{-1}$. This allowed producing enough resolution in mass and space to study both the DM and baryonic properties of the large-scale structures of the Universe, specifically the gravitationally bound galaxy clusters, in this work. For the hydrodynamic setup, an ideal gas equation of state, with $\gamma = 5/3$, has been used for the baryons. Radiative cooling and star formation feedback physics were also taken into account during the simulations (for more details, see \cite{Paul2017}).

\textbf{Simulated data sample:} With the above-said model, two realisations of parts of our Universe were produced, each of co-moving volume ($128$~Mpc\; $h^{-1}$)$^3$. Each simulation has been initialised at redshift, $ z = 60$ using the transfer function of \cite{Eisenstein1999} and evolved till the present epoch, i.e, $ z =0$, with 48 snapshots taken at different redshift intervals. In this work, we have used only 11 snapshots taken at equal intervals of $\Delta z=0.02$, in between redshift 0.2 to 0. To proceed further with the proposed algorithm, for each snapshot, the particles and the above-mentioned associated properties of the particles are recorded in a separate file. During the analysis, we worked only within the highly resolved central ($32$ Mpc $h^{-1}$)$^3$ co-moving volume of two realisations.

\subsubsection{\label{Sim_MITRO_algo}MITRO Algorithm for simulated data}
While finding DM halos from DM particles of a cosmological simulation, such as ours, using the MITRO code, it follows the below-mentioned basic algorithm 

\textbf{Step 1:} Defines the arm-length, $L_{i}$, to each particle following Eq.~\ref{eq:eq5} and appropriate parameters, i.e., $\Delta$, in general $\Delta=200$

\textbf{Step 2:} After defining the arm-length, it searches for friends of the particles in the volume of interest. Two particles, $i$ and $j$ are flagged as friends if the Euclidean separation between them ($D_{ij}$) is smaller than the sum of their arm-lengths; i.e., $D_{ij} \leqslant L_{i} + L_{j}$. Likewise, the search for the friends-of-friends is done to achieve the final configuration of the halos.

\textbf{Step 3:} Unbinding process (described in Sec~\ref{sec:unbinding}) is then applied to each MITRO-identified halo when required. It is an optional step which can be switched on or off depending on the science problem that we address.

\textbf{Step 4:} Finally, it computes the halo properties defined in section~\ref{basic_halo_properties}.

We put an additional restriction on particle numbers to flag an identified halo as the real halo (in both cases, i.e., before and after running the unbinding algorithm). We list only those halos having a minimum number of particles, $N_{min}=100$, decided upon the highest particle mass resolution of $\sim10^{10}$ M$_{\odot}$ in our simulation so that we obtain the halos with mass, $M\ge10^{12}$ M$_{\odot}$.

\subsubsection{\label{over-density_ratio}Over-density ratio of MITRO-halos and the analysis of unbinding process}
As we see in section~\ref{Compactness}, our proposed algorithm is capable of constraining the over-densities to a great extent, especially for the halos after running the unbinding algorithm. From the discussion in section~\ref{armlen}, it appears that to achieve a theoretical lower bound to the over-density for the searched halos, $\wp_{halo}$, one needs to keep in mind that an arm-length, $L_{arm}$ corresponds to an assumed over-density of $\Delta_{arm}$, would result in an approximate over-density lower bound of halos, $\wp_{halo} \geq \Delta_{arm}/2$. To statistically check the veracity of the above-stated feature of the MITRO algorithm, we take a practical example with $\Delta_{arm} = 200$, applied to the above-discussed simulated data set. Here, we should mention that for further analysis in this paper, we created a DM halo list following the algorithm illustrated in section~\ref{Sim_MITRO_algo} and taking all the identified halos with mass $\geq10^{12}$ M$_{\odot}$ (MITRO-200 hereafter).

\begin{figure}
    \includegraphics[width=\columnwidth]{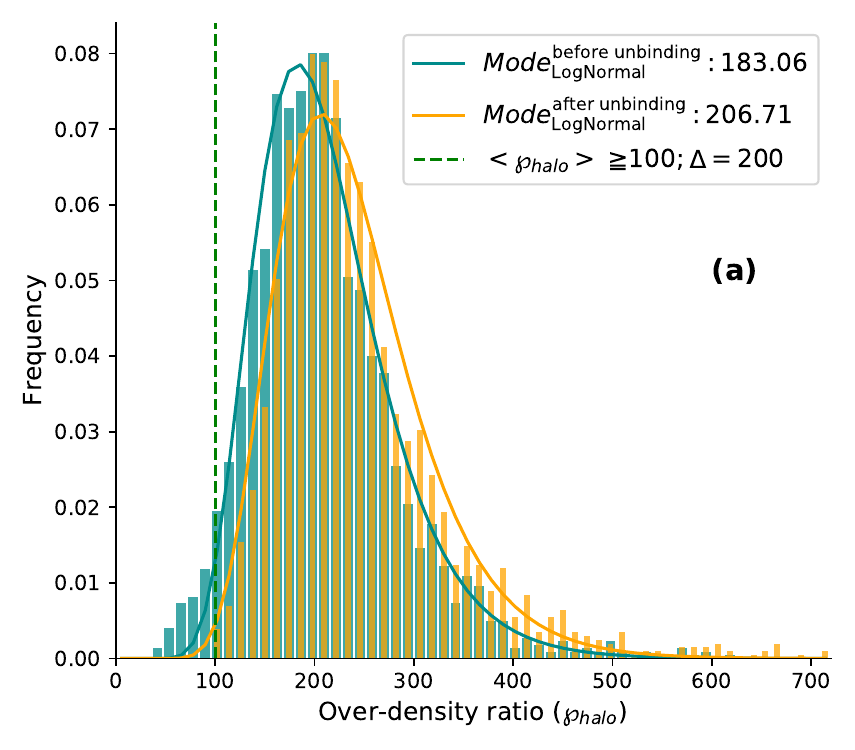}
	\includegraphics[width=\columnwidth]{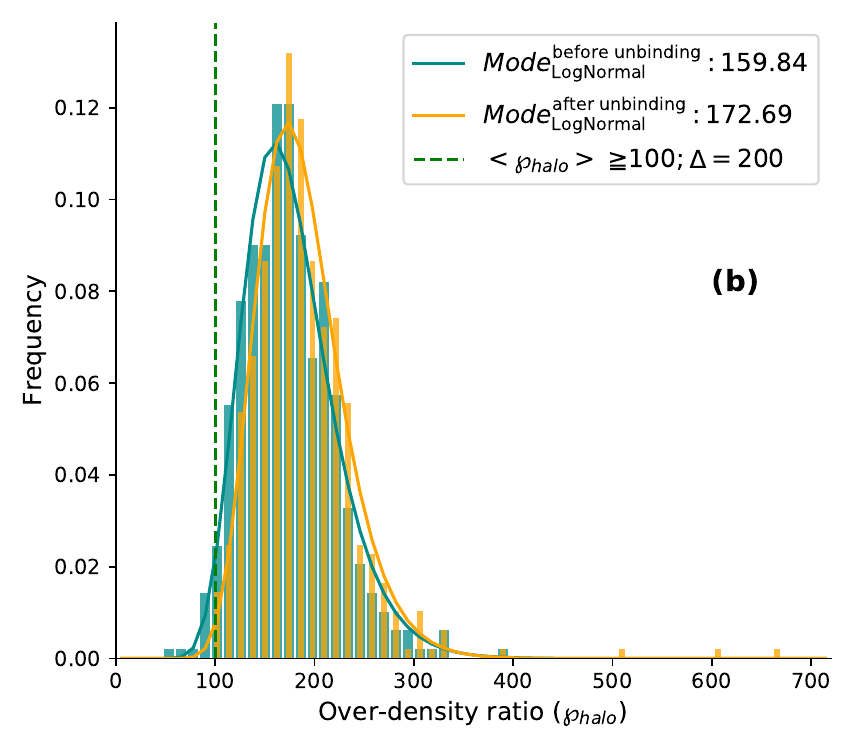}
    \caption{Over-density ratio of identified halos for a search over-density of $\Delta_{arm} = 200$. \textbf{Panel~a}: Histogram plot of $\wp_{halo}$ for all MITRO clusters having Mass $\geq 10^{12}$ M$_{\odot}$ \textbf{Panel~b}: the same for all MITRO clusters having Mass $\geq 10^{13}$ M$_{\odot}$)}
    \label{fig:Delta200}
\end{figure}

Figure~\ref{fig:Delta200} shows the distribution of over-density of the MITRO halos with the arm length corresponding to $\Delta = 200$ ($\wp_{halo}-dist$, hereafter). Figure~\ref{fig:Delta200}a depicts MITRO-identified halos with $M_{halo} \geq 10^{12}$  M$_{\odot}$, whereas, the lower panel (Fig.~\ref{fig:Delta200}b) represents the halos with $M_{halo} \geq 10^{13}$ M$_{\odot}$. In both the panels, dark-cyan colour has been assigned to halos identified before running the unbinding process and orange depicts the halos after the removal of unbound particles from the initially identified halos. The $\wp_{halo}-dist$ in Fig.~\ref{fig:Delta200}a~\&~\ref{fig:Delta200}b shows that the halos after going through the unbinding process strictly follow the expected lower bound, i.e., $\wp_{halo} \geq 100$, with 99.6\% of halos falling above the over-density lower limit. Whereas, before the removal of unbound particles, slightly more than 96\% of the identified halos falls above this determined lower bound of over-density. We further notice that the maximum probable over-density ratio in $\wp_{halo}-dist$ is roughly greater than the average of the lower-bound and the actual assumed over-density threshold used for defining the arm-length, i.e., $\gtrsim150$.

While, the $\wp_{halo}-dist$ for $M_{halo} \geq 10^{12}$ M$_{\odot}$ (Fig.~\ref{fig:Delta200}a), show a significant peak shift as well as a reduction in peak value between the distribution before and after the removal of unbound particles, the same is not so prominent in the distribution with $M_{halo} \geq 10^{13}$ M$_{\odot}$, as shown in Figure~\ref{fig:Delta200}b\footnote{A slight increment in the peak of distribution after the unbinding process is due to the increment in the compactness of low mass objects ($10^{13}$ M$_{\odot}\leq M_{\rm{halo}} \leq 5 \times 10^{13}$ M$_{\odot}$). The unbinding was found to bring the over-density ratio close to the searched over-density threshold ($\Delta_{arm} = 200$).}. This actually suggests that many low-mass objects, $10^{12}$ M$_{\odot}\leq M_{\rm{halo}} \leq 10^{13}$ M$_{\odot}$, are largely affected by the unbinding process\footnote{A few halos in the mass range $10^{12}$ M$_{\odot} \leq M_{\rm{halo}} \leq 5 \times 10^{12}$ M$_{\odot}$ are removed as they fail to qualify the minimum particle condition for a halo, i.e., $N_{min} = 100$ after losing particles due to unbinding process.}, indicating that they are very unstable and are possibly non-virialized, as also suggested by \cite{Paul2017}. This may also indicate an environment with shallower gravitational potential, such as filaments, easily allowing them to break their spherical geometry by any minor dynamical events (Gupta $\&$ Paul in prep.). Moreover, the large difference in maxima values of $\wp_{halo}-dist$ seen in Figure.~\ref{fig:Delta200}a for low and high mass halos makes it apparent that a significant fraction of the stable and gravitationally bound low-mass systems are compact in nature as further discussed in section~\ref{compactness_result}.

\subsubsection{\label{Position_offset}Position off-sets in MITRO halos}
For studying halo position off-sets, the local peak matter density of the configured halos, $\textbf{r}_{SO}$, have been taken as the reference position for the halos. It has been noticed that, under the smooth radial-density profile fits, the halo finders that work with the centre at the highest local density are more accurate than the one that computes the centres by taking the position average (simple or mass-weighted) of all the halo-fellow particles, such as in FoF or any position-space based halo finders (see \citet{Gao2006,Knebe2011}). Here, we demonstrate the accuracy of the MITRO-halo centres by defining the position off-sets as 
\begin{eqnarray}\label{eq:offset}
    \delta r_{com} = |\textbf{r}_{com} - \textbf{r}_{SO}|\;\;\; ; \;\;\;
    \delta r_{main} = |\textbf{r}_{main} - \textbf{r}_{SO}|
\end{eqnarray}

\begin{figure}
	\includegraphics[width=\columnwidth]{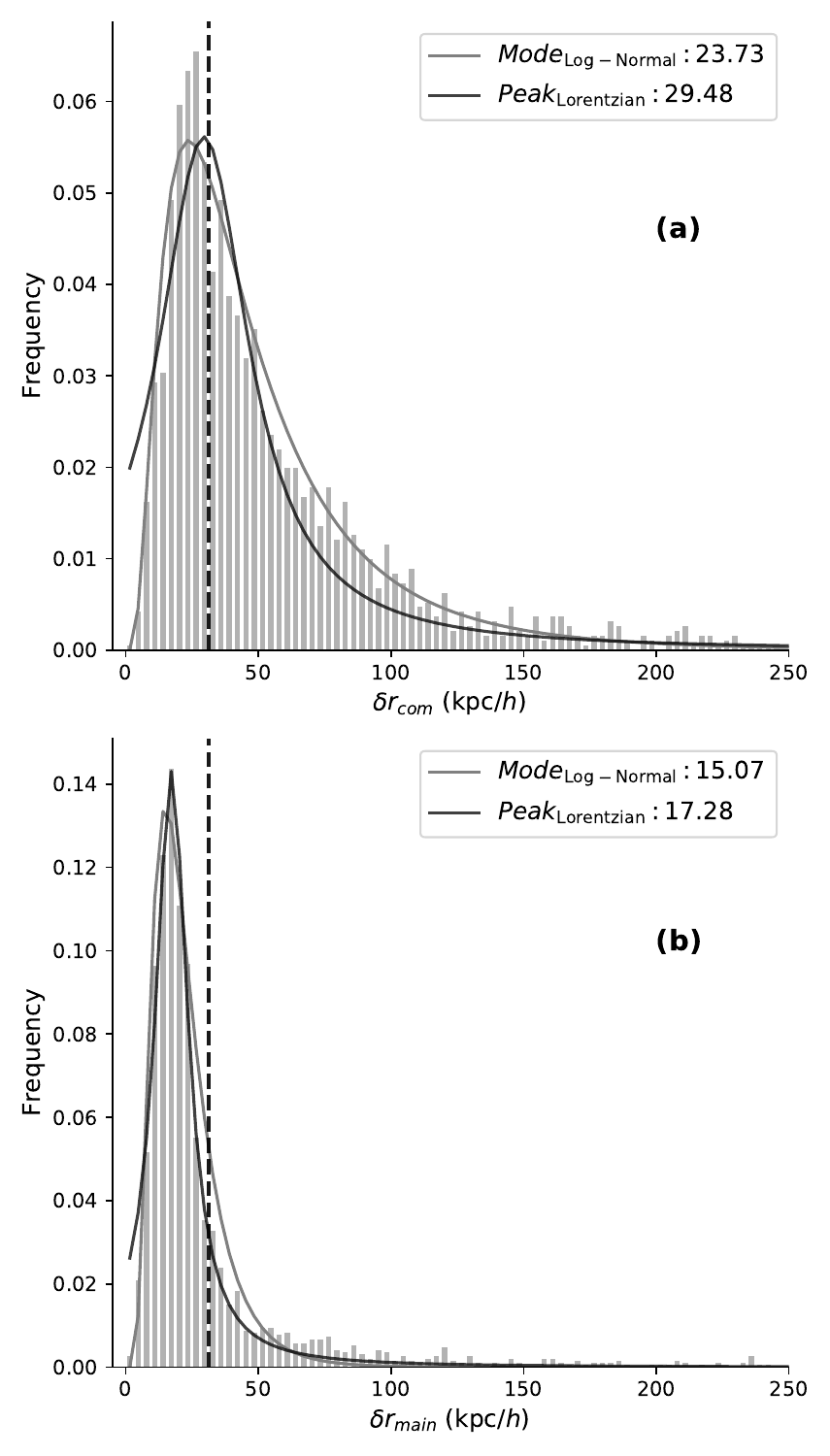}
    \caption{Halo position offsets between the mass-weighted average of positions of all halo particles, $\textbf{r}_{com}$, and the highest local density in the configured halo, $\textbf{r}_{SO}$, in \textbf{Panel a}, and between highest gravitational potential energy particle position, $\textbf{r}_{main}$ and $\textbf{r}_{SO}$ (\textbf{bottom Panel b}). The dashed black line represents the peak grid resolution of the simulations.} 
    \label{fig:C_offset}
\end{figure}

To search the local-density peak in the configured halos, we first create a particle density field with a uniform-grid structure in the halo-defined region, having the grid positions the same as given by ENZO snapshot outputs. The particle mass density field is produced using \textit{Triangular Shaped Cloud (TSC)} interpolation method (\cite{Hockney1988}), in which the mass of the particle will be distributed in 27 cells based on the spatial location of the particle in the grid-cells. Finally, the maximum density location is searched within the halo-defined region and is referred to as $\textbf{r}_{SO}$. Although we implemented AMR for our simulations, a uniform grid cell structure is created at the highest resolution (i.e., $31.25$~kpc/$h$) of our simulation during the reconstruction of the matter density field. However, it should be noted that such a crude process may affect the density field if any relatively high mass particle(s) resides at the periphery ($\gtrsim r_{200}$) of the configured halo.

The distribution of centre offsets computed following the Eq.~\ref{eq:offset} is plotted in Fig.~\ref{fig:C_offset}a~and~\ref{fig:C_offset}b. It is evident from the plots that MITRO computed centres (both ${r}_{com}$ and $r_{lgp}$) are precisely closed to the highest local density position, as the peak of centre-offset distributions is found to be well within the size of the highest resolution cell (i.e., $31.25$~kpc/$h$) in our simulations. The halo positions (${r}_{com}$) calculated based on the mass-weighted position average have a broader offset distribution than the one calculated based on the lowest gravity point (${r}_{lgp}$). The peak of the distribution of the former ($\delta r_{com}$) is also found much away compared to the latter ($\delta r_{lgp}$). Furthermore, while 73\% of the position off-set with respect to LGP centres (i.e., $\delta r_{lgp}$) are found to be within the highest grid-resolution of the simulation, i.e., $31.25$~kpc/$h$, only 37\% of position off-sets for CoM centres (i.e., $\delta r_{com}$) come within this resolution limit. All these shows, ${r}_{lgp}$ centres are more accurate and therefore used as the preferred definition for halo position in this work.

\subsubsection{\label{compactness_result}Compactness of MITRO-halos}
\begin{figure}
    \includegraphics[width=\columnwidth]{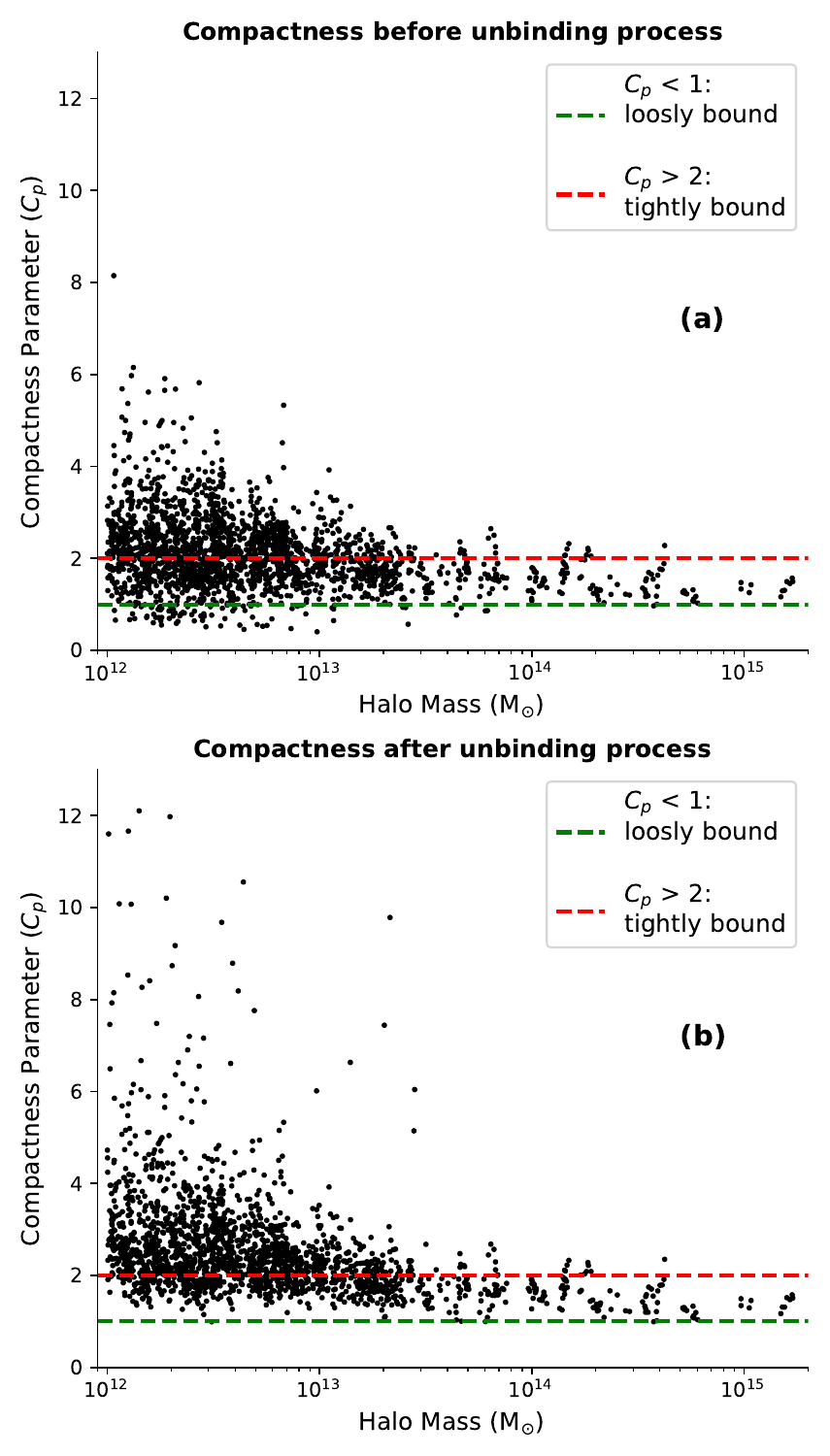}
	
    \caption{The scatter plot of compactness parameter, i.e., $C_{p}$, for formed halos with $\Delta_{arm} = 200$ as a decisive factor to assign the arm-length to the particles in search of member particles, for all MITRO halos having mass $\geq 10^{12} M_{\odot}$: without initialising the unbinding process (\textbf{Panel a}) and after the unbinding process (\textbf{Panel b}). The green dashed line corresponds to the unit $C_{p}$ below which halos are considered to be the loosely bound structures whereas the red dashed line corresponds to $C_{p}=2$ above which halos are tightly bound systems with respective over-density threshold in which the halos were searched.}
    \label{fig:Compactness_delta200}
\end{figure}

`Compactness' of a halo is defined in section~\ref{Compactness} as $C_p=\wp_{halo}/\wp_{ref}$, and it is a measure of how closely the halo-member elements are packed in a halo. By definition, unit compactness indicates a simply packed system, whereas the higher values point towards a tighter packing or overlapping of the arms of the halo elements. Values lower than unity are indicative of a loosely bound system. Figure~\ref{fig:Compactness_delta200}a~and~\ref{fig:Compactness_delta200}b exhibit a scatter plot of $C_{p}$ values of halos for both, without and with the enacted unbinding process, respectively. We mark $C_p \geq 2$ (red dashed line) as the limit for the bound too tightly bound systems, while, $C_p \leq 1$ (green dashed line) as the limit for marginally bound to loosely bound systems. It can be seen that with the chosen arm-length corresponding to $\Delta = 200$, the high mass systems ($M_{halo} \geq 5 \times 10^{13}$ M$_{\odot}$) have compactness within the range $C_p \sim$ 1 to 2, whereas, the compactness of low mass systems are spread over a wide range of values, $C_{p} \approx$ 1 to 6. It is also evident from the plots that the unbinding process filters out the larger number of loosely bound low-mass systems compared to that of high-mass systems, as also discussed in section~\ref{over-density_ratio}. We should mention here that a few low-mass halos exit the MITRO-halo list while the unbinding process was enacted as they could not meet the minimum particle condition ($N_{min} \geq 100$; see section~\ref{Sim_MITRO_algo}) any more, as during the unbinding it may have lost a few member particles. By explicit inspection, it is also found that some of the low mass halos lost their peripheral or satellite particles that resulted in a marginal decrement in their masses as well as in their extents, leading to enhancement of the mean density of the halos and consequently the increment of $C_{p}$ values. This can be easily observed in halos with mass $M_{halo} < 5 \times 10^{13}$  M$_{\odot}$ in figure~\ref{fig:Compactness_delta200}b.

Furthermore, if we were to understand the situation in which a halo become highly compact, we need to understand how the halos are formed. The hierarchical clustering model says, these structures continuously accrete material from the surroundings. Therefore, if the matter is available in the surroundings, the member particle distribution will fall gradually towards the outskirts and cuspiness in the halo will be less. While, lack of surrounding materials may lead to a cuspy halo because of the presence of a dominant core, leading to high $C_p$ values. In cosmic web structures, a lack of material can be foreseen as the void or a low-density region such as the cosmic filaments. Therefore, $C_{p}$ values can be a good indicator of the environment of the halos in which they are evolving. Here, we should clarify that it is not feasible to draw a general conclusion on the specific surrounding environment based on $C_p$ values only, mainly because of its intrinsic dependence on $\wp_{ref}$ which is controlled by the chosen value of $\Delta_{arm}$ for assigning the arm-length to each particle. Nevertheless, in general, a higher value of compactness parameter $C_{p}$ would mean the lower surrounding matter density for the halo with respect to the searched over-density threshold, as well as, it reveals the cuspiness of the identified halos. This roughly allows us to derive the type of environment in which the halo is evolving, and when the search value of over-density is known.

\subsection{\label{Observation_section}Finding supercluster scale structures in large redshift surveys using MITRO algorithm}
Mapping the Universe at large scales is best done with large galaxy surveys (i.e., 2dFGRS, SDSS, 2MASS, WISE etc.). Various observatories have done or are currently doing sky surveys with different science goals. However, in general, they collect the basic information, i.e., the photon counts of the sky patches in different optical colour bands. This is collectively processed in a database after filtering the objects and their associated parameters; basically, object type, colour bands magnitude, its position on the celestial sphere etc. The most crucial parameter that helps unfold the 3D structure of the Universe is the redshift information. A few of the surveys provide either photometric or spectroscopic redshift to infer the line of sight distance -- such surveys are called the galaxy redshift surveys \cite{Davis1982, Geller1989}. Though galaxy surveys are done at various frequency bands (e.g., X-ray, radio, optical etc.), among them the largest and deepest (in redshift) coverage is provided by the optical surveys, e.g. SDSS \cite{Stoughton2002, Ahumada2020}. Using the 2.5-meter Sloan Telescope \cite{Gunn2006} operated at the Apache Point Observatory, the Sloan Digital Sky Survey (SDSS) has provided by far the largest map of the Universe ($\sim$ 14,555 deg$^{2}$ sky coverage, and star-forming galaxy's redshift depth $\sim 0.8$) by constantly surveying the sky for over the last two decades \cite{Abdurrouf2022}. This large galaxy map of the Universe provides critical cosmological information on the key physical processes that govern the evolution of the Universe through its specific clustering patterns. To produce the clustering pattern from the raw galaxy survey data, one needs to identify the grouping of galaxies as precisely as possible as we have elaborated on this issue in sec.~\ref{intro}.
 
Here, we discuss the application of the MITRO algorithm to real observational data. We chose the Sloan Digital Sky Survey (SDSS-III DR12) galaxy data \cite{Stoughton2002, Alam2015} in the region of the most massive supercluster of galaxies, SARASWATI supercluster \cite{Bagchi2017} for our study, primarily to compare our results with the one given in the literature and understand how robust is our algorithm in the context of identifying most massive superclusters and its properties.

\subsubsection{\label{SDSS_sample}Data Sample}
We obtained the galaxy data from the SDSS data archive using the same constraint as \citet{Bagchi2017} (hereafter JB17) in the field of the SARASWATI supercluster to maintain consistency in data selection criteria. This would allow us to compare the results reported in JB17 with ours, which uses the proposed MITRO algorithm in a consistent manner. We queried (SQL-query using CAS-Job submission portal) the spectroscopic data from three different programs; `LEGACY', `BOSS', and `SOUTHERN'; of the SDSS-III DR12 database for galaxies having redshift errors $<1\%$, and clean photometry within the wedge of R.A. and Dec. range $336^{\circ} \leqslant {\rm{RA}} \leqslant 16^{\circ}$, $-1.25^{\circ} \leqslant {\rm{Dec}}. \leqslant +1.25^{\circ}$, and spectroscopic redshift range $0.23 \leqslant z \leqslant 0.33$. The data from all three programs were combined together to give out a total of 3136 galaxies. Furthermore, we also obtained the SDSS apparent magnitude for each of the galaxies as an additional physical parameter required by the MITRO algorithm (see Sec. ~\ref{Survey_arm} for more details). We adopt the extinction corrected $model$ apparent magnitude calculated from the best of two fits (a de-Vaucouleurs and an exponential luminosity profile) in the r-band. Here, to mention, we preferred the $model$ apparent magnitude in order to measure the unbiased colours of the galaxies, since their flux is measured in the equivalent aperture for all the SDSS bands, which makes the galaxy colour an aperture-independent parameter. The absolute magnitudes of galaxies are $k$-corrected using \textit{K-CORRECT} (version 4.2 python) software \cite[][hereafter BR07]{Blanton2007} for the rest frame at $z=0$, and evolution corrected $E_{b}(z) = E_{0_{b}}z$ using $E_{0_{b}} = \{2.3,\; 1.6,\; 1.3,\; 1.1,\; 1.0\}$ in $b = \{u,g,r,i,z\}$ \cite[][hereafter B03]{Bell2003}.

\begin{figure*}
   \includegraphics[width=\textwidth]{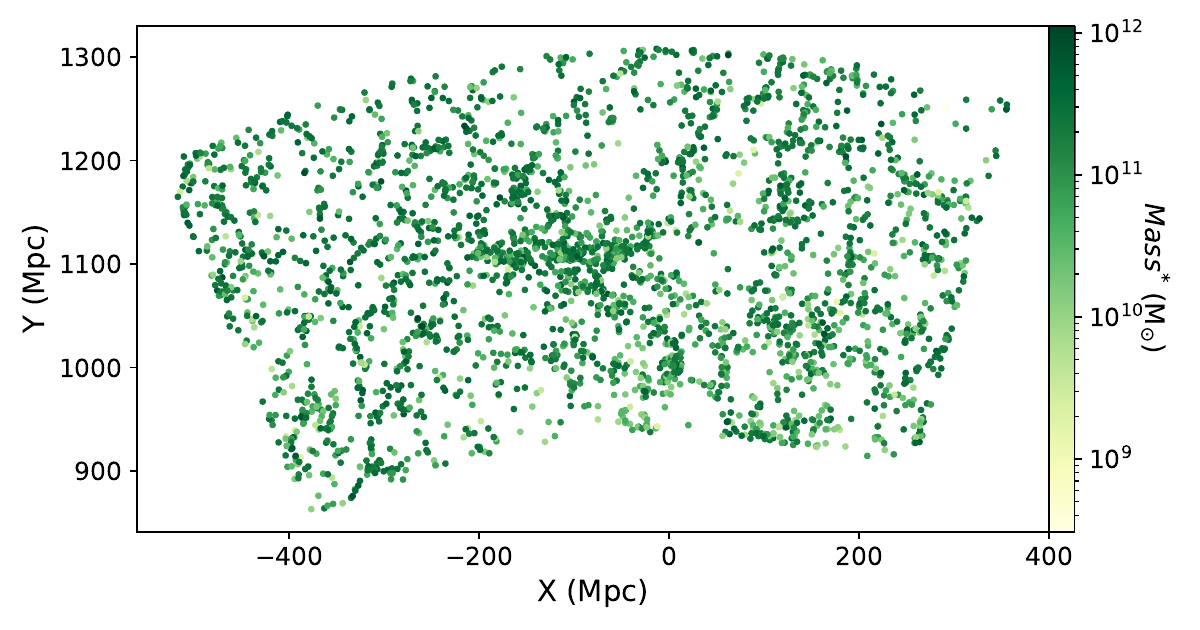}
    \caption{\label{fig:SDSS_Galaxy}Selected wedge of 3136 galaxies plotted in co-moving coordinates, projected on the two-dimensional $X-Y$ plane. The declination along the $Z$-axis has been suppressed.}
\end{figure*}

For further calculations, required to run our group finding algorithm, we produced the co-moving Cartesian co-ordinates of each galaxy in the above-discussed (flux-limited) sample using the relations;
\begin{align}
    x_{i} &= R_{i}\; cos(\delta_{i})\; sin(\theta_{i}) \nonumber \\
    y_{i} &= R_{i}\; cos(\delta_{i})\; cos(\theta_{i}) \\
    z_{i} &= R_{i}\; sin(\delta_{i}) \nonumber
\end{align}
where $R_{i}$ is the co-moving distance of the $i^{th}$ galaxy, and $\theta_{i}\;\&\;\delta_{i}$ are their respective right ascension and declination. For computing various quantities, we adopted the cosmological parameters based on five-year $WMAP$ results \cite{Komatsu2009}; $\Omega_{M} = 0.279,\; \Omega_{\Lambda} = 0.721,\; \Omega_{R} = 8.493\times10^{-5}$ and Hubble parameter, $H_{0} = 70.1$ km s$^{-1}$ Mpc$^{-1}$ (same as followed by JB17, however, we verified that the adopted value of $\Omega_{R}$ does not affect the final results). The galaxy distribution within the selected wedge in the translated co-moving Cartesian coordinate space has been shown in Fig~\ref{fig:SDSS_Galaxy}.

\subsubsection{\label{Survey_arm}Applying MITRO algorithm on the survey data}
Before we start finding the over-dense regions in the survey data, we first made the data compatible with the structure of the proposed algorithm that is originally written on the basis of the simulated data. Since the principal basis of the proposed algorithm is to assign the physically motivated arm-length to each and every element (here the galaxies) in the search volume, as discussed in sec.~\ref{model}, which we achieved by correlating the arm-length with the mass of each galaxy. The choice of mass as the parameter for assigning the arm-length is robust as gravitational force is the dominant of all forces acting at large scales, essential for building the bound structures in the Universe. However, unlike the particles in simulated data, no direct mass information is available for the galaxies in the SDSS data sample, and therefore we need to estimate the masses of galaxies in the selected SDSS wedge.

It is well known that the mass composition of the galaxies is largely dominated by dark matter along with a non-negligible visible baryonic mass. It is the total mass (i.e. Baryon+DM) of individual galaxies that determines the gravitational field of influence, and is essential information required by the MITRO algorithm for computing the arm-length parameter. Although, as a first approximation, the stellar-to-halo mass relation \cite{Vale2004, Zheng2007, Yang2012, Puebla2017, Girelli2020} would provide a better proxy to compute the total mass of galaxies, this relation does not always provide the mass of local halo of individual galaxies, rather, it usually provides the mass of cluster DM halo in which the galaxy and their companion galaxies are evolving, making it unfit for our case. Therefore, as an alternative, we first estimate the colour-derived stellar mass, $M_{*}$ of each galaxy in the selected wedge by adopting the correlation, discussed in \citet{Bell2003} (hereafter B03), between the stellar mass-to-light ratio and colour of the galaxy. \citet{Bell2003} report that at redshift, $z = 0$, $log(M_{*}/L_{r}) = 1.097 \; ^{0}(g-r) - 0.306 - $offset, where the $M_{*}/L_{r}$ ratio is in solar units. We took the effective offset of $0.10$ corresponding to the \citet{Kroupa2001} initial mass function (IMF) \cite{Borch2006}. Stellar masses $M_{*}$ were then obtained by multiplying $M_{*}/L_{r}$ to the rest-frame r-band luminosity $L_{r}$ (which is not evolution corrected). The final relation that we obtained for the stellar masses of the galaxies is,
\begin{equation}
log \left[ \frac{M_{*}}{\text{M}_{\odot}} \right] = 1.097\; ^{0}(g-r) \; -\;0.406 \;+\; 0.4 \left[M_{\odot,r}\;-\;^{0}M_{r} \right]
\label{eq:grcolorMass}
\end{equation}
where $M_{\odot,r} = 4.64$ is the r-band absolute magnitude of the Sun in the AB system (BR07), and $^{0}(g-r)$ and $^{0}M_{r}$ are the $(g-r)$ colour and r-band k-corrected absolute magnitude to redshift $z = 0$, respectively. We computed the $K+E$ corrected absolute magnitude in band-pass $b$ for each galaxy using;
\begin{equation}
^{0}M_{b} = m_{b} + {\Delta b_{AB}} - DM(z) - K_{b} - E_{b}(z)
\end{equation}
where $\Delta b_{AB} = \{-0.036,\; 0.012,\; 0.010,\; 0.028,\; 0.040\}$ for $b = \{u,g,r,i,z\}$ \cite{Yang2007}, is the latest zero-point correction for converting the apparent magnitudes ($m_{b}$) of SDSS to the AB system. $DM(z) = 5log[D_{L}/\text{Mpc}] + 25$ is the bolometric distance modulus corresponding to the luminosity distance $D_{L}$ in $WMAP$ cosmology (given in section ~\ref{SDSS_sample}). $K_{b}$ and $E_{b}(z)$ is the k-correction and evolution correction to $z=0$. Out of 3136 galaxies in the list, $^0(g-r)$ colour-derived-stellar mass of two galaxies was found to be unrealistically high, i.e., $>10^{14}$ M$_{\odot}$, plausibly due to incorrect colour magnitude in the specific band. For them, we obtained the stellar mass from $^0(r-i)$ colour using the relation (refer Table 7 of B03);
\begin{equation}
log \left[ \frac{M_{*}}{\text{M}_{\odot}} \right] = 1.431\; ^{0}(r-i) \; -\;0.122 \;+\; 0.4 \left[M_{\odot,r}\;-\;^{0}M_{r} \right]
\end{equation}
Finally, we obtained stellar mass for all the galaxies that came out to be in the range $10^{9} - 10^{12}$ M$_{\odot}$, and plotted them as colour-map in Fig~\ref{fig:SDSS_Galaxy}. Once we compute the stellar masses, the next task was to find the dark matter mass associated with each of the galaxies which is not a very straightforward task to do. Furthermore, in the spectroscopic galaxy survey (SDSS-DR12) that we took for this study (same as the JB17), the number density of the galaxies is quite low in the examined volume and therefore would introduce a significant missing galaxy problem. Since the proposed algorithm defines its arm-length on the basis of the over-density ratio (see sec.~\ref{armlen}), the corresponding arm-length associated with the stellar mass would be unphysically shorter in sample wedge where galaxy number density is lower because of missing galaxies. In the following section, we would come up with an approximate solution from the basic principles of cosmology that would significantly mitigate this problem.

\subsubsection{\label{missingmatter}Accounting the missing matter (DM and the missing galaxies)}
Since the MITRO algorithm defines its arm-length on the basis of the over-density ratio parameter (see sec.~\ref{armlen}) while searching the structures, having information on all materials present in the volume of interest is of paramount importance. However, unlike the simulations, where the complete information of matter (DM and baryons) and energy densities are readily available; optical galaxy surveys can only provide the information of stars and galaxies i.e., merely a fraction of baryonic matter density of the Universe. Information on the intergalactic medium or the DM halos is completely missing. Although the indirect evidence of DM can be obtained from the gravitational lensing effects, it needs dedicated observations and analysis of the fields and is therefore out of the scope of this work. Furthermore, galaxy surveys are intrinsically flux-limited depending on the depth of the observations. Due to the dimming effect \cite{Tolam1930}, objects with the same luminosity but placed at a higher redshift may remain undetected, while at a lower redshift, they may get detected if they are above the sensitivity limit of the instrument(s). So, many faint galaxies present at higher redshifts will remain undetected if are below the sensitivity of the instrument of a galaxy survey, causing missing matter problems in the volume of interest.

To take account of this missing matter (due to undetected galaxies and DM content as well as intergalactic medium), we provide a very simple solution from the basic principles of cosmology. We began by dividing the $3$-D redshift-space of the sample in n parts (n=10 for our analysis) along the line-of-sight keeping the co-moving volume enclosed by each part to be equal and greater than $(100~ {\rm{Mpc}})^{3}$. This particular volume is chosen to retain the homogeneity scale of the Universe, the scale at which the statistical properties of the Universe have transnational invariance. \citet{Bharadwaj1999,Ntelis2017} have reported this scale to be $\sim63 h^{-1}$~Mpc $\approx 90$~Mpc, at which the observables, i.e., the mean counts of neighbouring points in the D-dimensional space and the fractal correlation dimension, approaches its homogeneous value. The cosmological mean-matter density would therefore account for all the material present in each of these sections (say, $\rho^{i}_{m_{c}}$ in $i^{th}$ section), whereas, the matter density measured using the colour-derived stellar mass of the galaxies (see sec.~\ref{Survey_arm}) in each section (say, $\rho^{i}_{m_{SDSS}}$ in $i^{th}$ section) will be much lower as it considers only some fraction of mass (i.e., baryonic) information luminous only in optical energy band. Therefore, the difference between these two densities may tentatively account for the missing matter in the sampled volume. To compensate for this mass, we simply distributed the unaccounted mass to each galaxy in the respective co-moving volumetric section. For a rationalised distribution, a weight factor ($w_{i}$) is computed for each $i^{th}$ section by taking the ratio of cosmological mean matter density at the mean-redshift of the co-moving volumetric section to the corresponding total stellar mass density; $w_{i} = \rho^{i}_{m_{c}} / \rho^{i}_{m_{SDSS}}$ and multiplied it to their individual stellar mass i.e., $w_{i} \times {M_{*}}_{ki}$, where ${M_{*}}_{ki}$ is the stellar mass of $k^{th}$ galaxy in $i^{th}$ volume sector. The missing matter from the surroundings thus has been distributed to each of the galaxies in the $i^{th}$ section in proportion to their stellar masses. The best part of this approach is, even if the number of galaxies in a field increases due to the higher sensitivity of any future surveys or if we include the SDSS photometric data along with our current spectroscopic sample (where available), the results would not get affected considerably (provided the increased number density of galaxies is uniformly distributed over the examined volume). Since the change in galaxy number density due to the availability of more galaxies in the field would modify the stellar mass density ($\rho^{i}_{m_{SDSS}}$) of the field, the corresponding weight factor will get adjusted accordingly. For the same reason, the issue of increasingly high missing galaxies in the deeper redshift fields due to the dimming effect and flux limitations, which usually needs extra attention while computing various clustering parameters such as the linking length (in FOF), would automatically get adjusted in our proposed MITRO algorithm (more details in sec.~\ref{SDSSarm}).

\subsubsection{\label{SDSSarm}Arm-length in RA-DEC plane and along the line-of-sight}
A great deal of discussion has been done in sec.~\ref{armlen} regarding the concept of arm-length in our proposed algorithm. To assign arm-length for individual galaxies in a field, the most crucial information is the total mass ($M$) of each galaxy, which, in the case of observational surveys, we attempted to obtain by the method described in sec.~\ref{missingmatter}. The arm-length of each galaxy has been determined by the Eq.~\ref{eq:eq5}, where $M$ is the effective mass of the galaxy, given as, $M_{k} = M_{*_{ki}} w_{i}$, and $\rho_{c}$ is the critical density of the Universe at the redshift of the galaxy. The other important parameter is the choice of $\Delta$ which decides the arm-length for the individual galaxies. As we mentioned earlier, while finding the local over-dense regions, $\Delta$ is the measure of the matter over-density with respect to which the structures are identified. Since we are interested in identifying the very large structures, specifically the galaxy superclusters, appropriately the over-density parameter has been chosen. Based on the adopted cosmology, to ensure sustainable gravitationally bound structures at the supercluster scale, the required bare minimum density contrast is suggested to be $\rho / \rho_{c} > 2.36$ \cite{Dunner2006}. Accordingly, we chose $\Delta = 2.36$ as a reasonable approximation for our further analysis. In the entire volume of interest (Sec.~\ref{SDSS_sample}), with $\Delta = 2.36$ and the other above-discussed parameters, we obtained the individual arm-length for the galaxies in the range of $680$~kpc to $10.8$~Mpc according to their weighted masses.

While this fixes the arm-length in R.A.-Dec. the plane, finding an arm-length along the line of sight needs careful consideration. The redshift (cosmological redshift) of the extra-galactic sources (here, galaxies) has been used to estimate their distances along the line-of-sight (LoS) which provides the de-projected position information from the inner surface of the celestial sphere. However, this de-projection of galaxy positions on the LoS is largely affected by redshift distortions (RSD). It is generally assumed that the observed redshift is the result of cosmological expansion only, however, in practice, the recession velocity of the objects also manifests the peculiar velocity induced by the gravity of neighbouring objects. This perturbs the cosmological redshifts of the objects and hence distorts the clustering pattern of galaxies in redshift space. On the small scales, this elongates the galaxy cluster along the LoS, known as the Finger of God effect \cite{Jackson1972}, whereas, it squashes the clusters at large scales, called as Kaiser effect \cite{Kaiser1987}. Although there are additional effects induced by general relativity, the gravitational redshift distortion \cite{McDonald2009} and Sachs-Wolfe effect \cite{Sachs1967}, the effects are quantitatively insignificant. Apart from these dynamical distortions to the redshift space, the difference between the adopted cosmology (while computing the line of sight distances to the galaxies, using their redshift information) and the true underlying cosmology of the Universe, artificially induces the geometrical distortions in the clustering of galaxies, known as Alcock-Paczynski (AP) effect \cite{AP1979}. There is no direct method to correct all these possible distortions in redshift space, at least to the observational data \cite{Zhu2017, Yu2019, Wang2020, Paillas2021}.

Here, we attempted to address this issue by borrowing an idea from the basic approach used for defining the linking length in traditional FoF. The traditional FoF-Group finders while searching clusters in galaxy surveys, adopted two linking lengths approach. The one that is along the LoS (in the radial direction) and the other one is transverse of it (on the 2-D projected sky-plane). Therefore, instead of searching for friends within the spherical region around the individual galaxy, one needs to do the same within an ellipsoidal region. However, in practice, it is done in a cylindrical region, as a cylindrical kernel is found to recover groups better than the ellipsoidal one \cite{Eke2004}. In the cylindrical kernel, the transverse linking length ($l_{\perp}$) is the radius of the cylinder and the length of the cylinder is taken as the linking length ($l_{\parallel}$) in the radial direction. Observationally, due to RSD effects, the co-moving LoS separations between two galaxies within the group appear larger than the projected separations. However, these linking lengths cannot be estimated directly from the galaxy survey data. Instead, they are determined from the analysis of mock galaxy catalogues from N-body simulations, where the true clusters are known. In literature, the typical ratio for the radial to transverse linking length (i.e., $\gamma_{gr} \equiv l_{\parallel}/l_{\perp}$), is found to be in the range between $5-20$ \cite{Tago2010, Tempel2014}, and are chosen as a fixed number for all galaxies in a particular study, depending on their science goals. Such an approach of fitting all galaxies with a single LoS arm-length (in MITRO) would go against our basic idea of finding linking parameters from the physical quantities reflected from the contribution of individual elements, as done for the transverse linking length. Also, an exercise to find the most suitable combination of \{$l_{\parallel},l_{\perp}$\} from any simulated mock galaxy catalogue to test the recovery of true simulated galaxy clusters is out of the scope of this study. For a straightforward application of the MITRO algorithm on galaxy redshift survey data, keeping the main essence of our proposed algorithm intact, we adopted a different but theoretically motivated way to compute the radial-to-transverse linking length ratio, $\gamma_{gr}$.

As we already discussed, the usual elongation of the linking parameter, along the LoS, depends on the motions of the galaxies inside the halo. Thus the general notion of the peculiar velocity of the galaxies inside the halos may provide a clue to overcoming the RSD effect. In a simplistic approach, we assumed each galaxy of the sample as a part of a halo, with halo mass $M_{h}$ and radius $r_{h}$, where, the virialization process would ensure the one-dimensional velocity dispersion \cite{Bryan1998} to be
\begin{eqnarray}
\sigma_{v,z} = \left( \frac{\Delta}{16} \right)^{\frac{1}{6}} \left( G M_{h} H_{z} \right)^{\frac{1}{3}}
\label{eqn:Bryan_veldisp}
\end{eqnarray}
where $\Delta$ is the mean density contrast relative to the critical density at the halo redshift, and $H_{z}$ is the redshift-dependent Hubble parameter. Using $\rho_{c} = \frac{3 H_{z}^2}{8 \pi G}$, $M_{h} = \frac{4\pi}{3} \Delta \rho_{c} r_{h}^{3}$, and with some elementary mathematics, Eq.~\ref{eqn:Bryan_veldisp} reduces to;
\begin{eqnarray}
\frac{\sigma_{v,z}}{r_{h}} =  \sqrt{\frac{2}{3} \pi G \Delta \rho_{c}}
\label{eqn:velocity_vs_radius}
\end{eqnarray}
Equation~\ref{eqn:velocity_vs_radius} expresses a velocity equivalent to the peculiar velocity of the constituent galaxies inside the virial radius of the halo. At the present redshift, for the adopted cosmology, $\Omega_{M} = 0.279, \Omega_{\Lambda} = 0.721$, the spherical collapse model provides the value of $\Delta \approx 100 $ for barely virialized clusters. This gives rise to;
\begin{eqnarray}
\frac{\sigma_{v}}{r} \approx 350~\rm{km\; s^{-1} Mpc^{-1}}
\end{eqnarray}

\begin{figure}
   \includegraphics[width=\columnwidth]{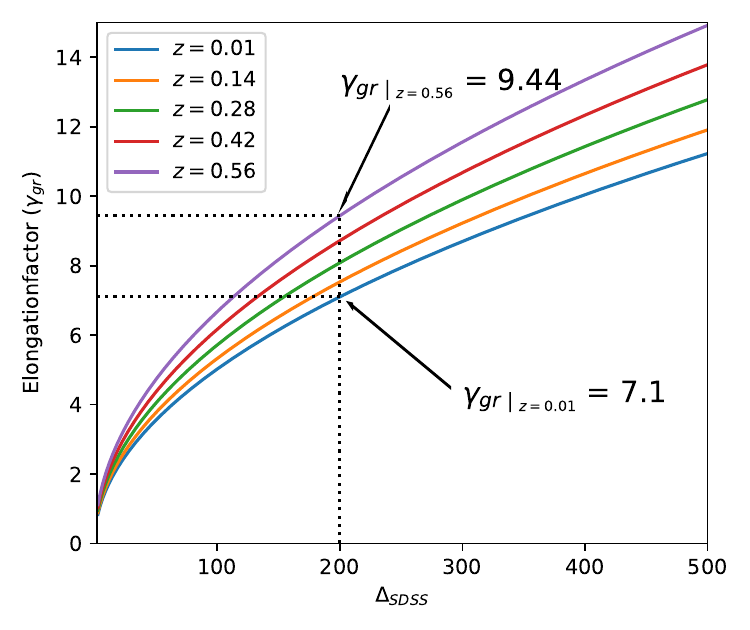}
    \caption{Variation of the elongation ratio with the chosen density threshold $\Delta$ while defining the arm-length to each galaxy. At higher redshift, the elongation factor increases for the definite choice of density threshold $\Delta$.}
    \label{fig:elongation_factor}
\end{figure}

Consequently, in a virialized structure of radius $r$, the peculiar motions of the constituent galaxies can stretch the structure to $1\sigma \sim \frac{3.5}{h}r$ along the line of sight. This makes the ratio of LoS to the transverse size of the object independent of $r$ and is $\frac{3.5}{h} \sim 5$ ($h$ = 0.701). Intriguingly, such a theoretically motivated number turns out to be quite consistent with the elongation factors $\sim 5-15$  (i.e., well overlapping within $1\sigma - 3\sigma$ range) adopted by various authors \cite[e.g.][]{Tago2010, Tempel2014} while finding groups using the traditional FoF method. Furthermore, we speculate that each galaxy in $\mathsf{S_{SDSS}}$ as a part of some pseudo galaxy cluster, therefore the maximum distortion it can introduce in the cosmological redshift is $\sim (1\sigma - 3\sigma)$ with respect to the searched over-density threshold. However, while clustering, instead of correcting the galaxy redshift space due to RSD, we compensated for it by introducing an elongation in the `arm-length' along the line of sight. We assumed the $1\sigma$ elongation factor while assigning the arm-length along the line-of-sight ($l_{\parallel}$) in our algorithm. Therefore, it boils down to the elongation ratio of $\gamma_{gr} = \frac{\sigma_{v}}{rH_{0}}$, where $H_{0}$ is the Hubble parameter. Since the critical density $\rho_{c}$ is redshift dependent, the over-density parameter, $\Delta$ (here, $\Delta_{\mathsf{SDSS}}$) and concomitantly the elongation in arm-length for individual galaxies would be different depending on their redshift. In the selected $\Delta_{\mathsf{SDSS}}$ (Sec.~\ref{SDSSarm}) and adopted cosmology with $H_{0} = 70.1 $km s$^{-1}$ Mpc$^{-1}$, the values of $\gamma_{gr}$ ranges between $0.86-0.9$. For a better illustration of the variation of the elongation factor with respect to the over-density parameter and redshift of the examined volume, we produced a plot shown in Fig.~\ref{fig:elongation_factor}. The plot highlights that for the usual virialized structures (with $\Delta=200$) formed within redshift $z=0.01-0.60$, the elongation factor ranges between 5-10. However, the structures at superclusters scales, whose over-density factor is usually much lower at about $\Delta=2-3$, result in an elongation factor of $0.77 -1.05$ indicating wall-like structures. This situation is quite similar to the Kaiser squashing effect, however, may not represent this as a general case.

\subsubsection{\label{SDSS_Mitro_algo} MITRO Algorithm applied to galaxy survey data}
The basic MITRO algorithm for finding large-scale bound structures is as described below:\\
\textbf{Step 1:} Compute the colour-derived stellar mass, $M_{*_{i}}$, of each galaxy in the sample (Eq.~\ref{eq:grcolorMass}).$\;\; M_{*_{i}} \;\; \forall \;\; i \in \mathsf{S_{SDSS}}$\\
\textbf{Step 2:} Estimate the weight factor, $w_{i}$, for each galaxy; to account for missing galaxies, IGM and DM in the survey sample (Sec.~\ref{missingmatter}).\\
\textbf{Step 3:} Define the arm-length, $L_{i}$, to each galaxy using Eq.~\ref{eq:eq5} and relevant parameters as discussed in Sec.~\ref{SDSSarm}.\\
\textbf{Step 4:} Estimate the radial-to-transverse link-length ratio, i.e., the elongation ratio $\gamma_{gr}$ (Sec.~\ref{SDSSarm})\\ 
\textbf{Step 5:} With the arm-length as the radius (transverse length) in $R.A.-Dec.$ plane, and line-of-sight (radial length, determine from $\gamma_{gr}$) as the depth, create cylindrical regions for each galaxy such that galaxy is placed at the middle of the axis of the cylinder.\\
\textbf{Step 6:} Now search for friends in the selected wedge - whenever two cylinders correspond to the pair of galaxies intersect each other, those pair of galaxies are considered as friends. And finally, search for the friends-of-friends to get the final configuration of the bound structures in examined sky volume, with a minimum of 10 galaxy constraints.\\
\textbf{Defining friend:} Two galaxies, $i$ and $j$, are friends if they satisfy the following conditions :
\begin{itemize}
    \item the angular separation ($\theta_{ij}$) between two galaxies at co-moving distances $R_{i}$ and $R_{j}$, and arm-length $L_{i}$ and $L_{j}$ respectively; 
    \begin{equation}
        \theta_{ij} \leqslant \frac{L_{i}}{R_{i}} + \frac{L_{j}}{R_{j}}
        \label{angle_friend}
    \end{equation}
    \item and LoS difference;
    \begin{equation}
        |R_{i}-R_{j}| \leqslant \gamma_{gr}^{i} L_{i} + \gamma_{gr}^{j} L_{j}
        \label{los_friend}
    \end{equation}
\end{itemize}

As a preliminary test, we implemented this modified MITRO algorithm on a simulated mock galaxy sample that roughly mimics the real observational data. Furthermore, before applying the proposed algorithm to real observational data set, we produced the halo-mass function of MITRO-identified structures and compared them with the standard theoretical halo-mass functions (see Appendix ~\ref{MITRO_on_Mock_Sample} for more details).

\subsubsection{MITRO identified supercluster scale structures in SDSS-DR12 sample}

\begin{figure*}
   \includegraphics[width=\textwidth]{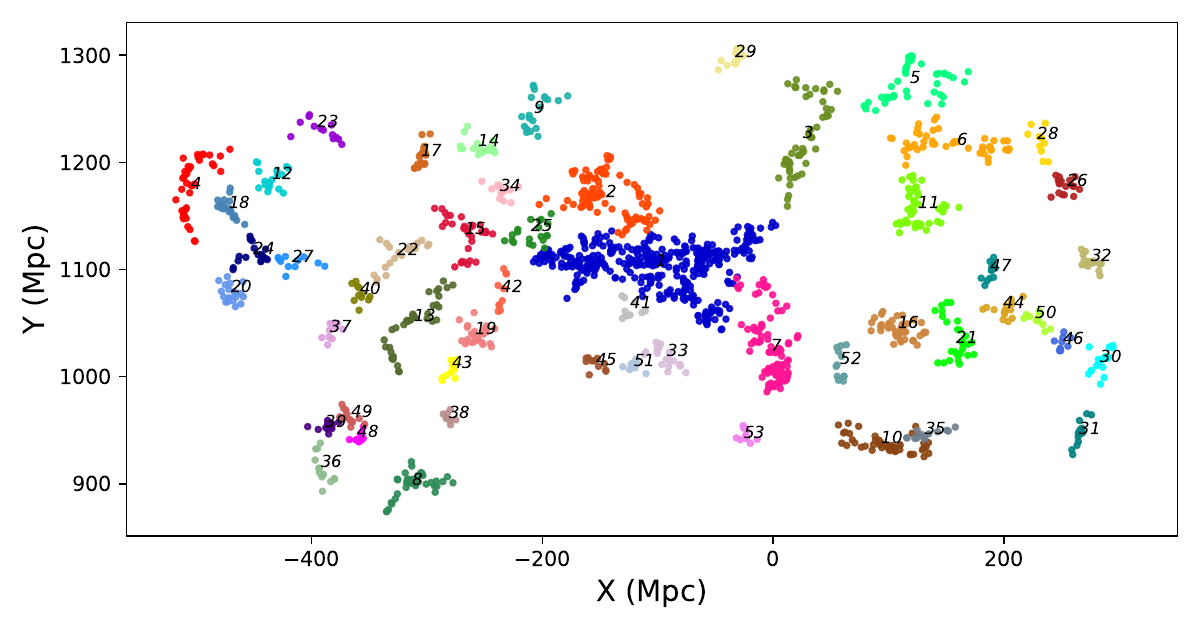}
   \caption{\label{All_SDSS_structures}Structures in $\mathsf{S_{SDSS}}$ using MITRO-algorithm having minimum 10 number of galaxies. Different colours indicate different structures with their unique structure IDs as plotted in co-moving coordinates, projected on the two-dimensional $X-Y$ plane. Declination along the $Z$-axis has been suppressed.}
\end{figure*}

Finally, we implemented our algorithm on galaxy survey data to identify supercluster scale structures. In order to search the over-dense regions in observational data, we introduced various additions and modifications to the data as well as to the algorithm, as discussed in the above sections. We choose the spectroscopic galaxy survey sky wedge from SDSS-DR12 in the field of Saraswati-supercluster by running relevant SQL queries (sec.~\ref{SDSS_sample}) as the data set. The choice of the over-density value is the most crucial parameter for defining an identified structure as supercluster \cite[see][]{Chon2015}, specifically to ensure that the structure remains gravitationally bound at present or will eventually collapse in the future. Since our principal aim is to identify supercluster scale structures, as discussed in Sec.~\ref{SDSSarm}, as a reasonable approximation derived from the spherical collapse model, we choose the over-density parameter to be $\Delta \equiv \Delta_{SDSS} = 2.36$ \cite{Dunner2006, Chon2015} for running our algorithm.

The selected SDSS data wedge consists of 3036 galaxies. It is intrinsically a flux-limited sample, and also not uniformly sampled in the redshift space. Furthermore, as only a fraction of baryonic matter is observed in the form of the galaxy in these surveys, most of the matter density is missing though out. To compensate for this surrounding missing matter, we assigned a weight ($w_{i}$) factor to each galaxy in the sample following the method described in section~\ref{missingmatter}. Eventually, the arm-length to each galaxy is defined using the weighted mass of each galaxy, i.e. $w_{i}M_{i}^{*}$, at a specified over-density and identified the over-dense structures in $\mathsf{S_{SDSS}}$ by running the MITRO clustering algorithm (sec.~\ref{SDSS_Mitro_algo}).

\begin{figure*}
   \includegraphics[width=\textwidth]{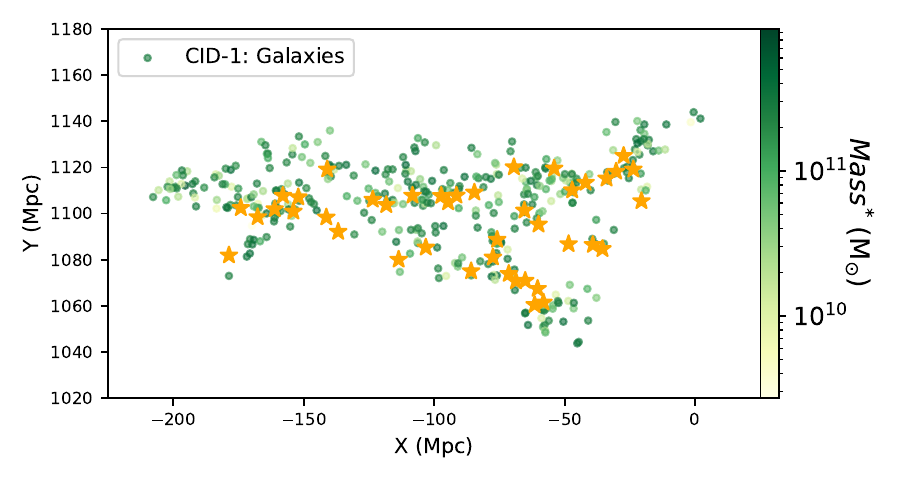}
   \includegraphics[width=\textwidth]{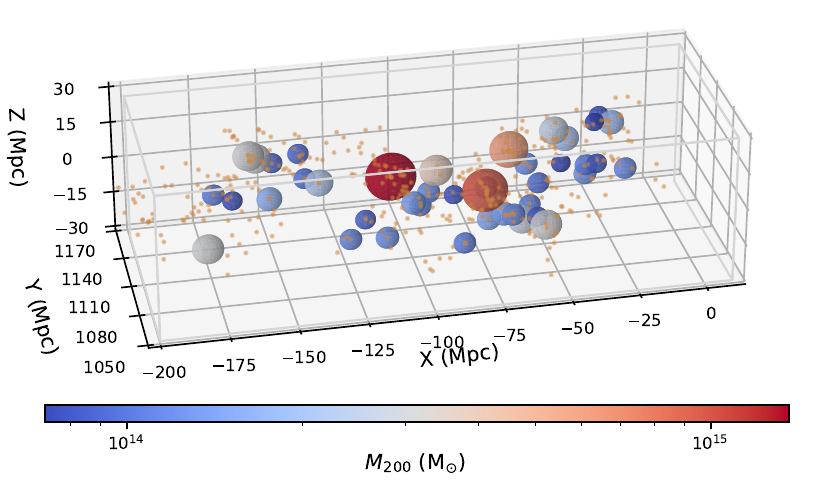}
   \caption{\label{Saraswati_cluster3d}\textbf{Upper Panal:} Saraswati-supercluster plotted in co-moving coordinates, projected on the two-dimensional $X-Y$ plane, spanning $\sim 200$ Mpc with 384 galaxies in it. Colours represent the stellar mass of fellow galaxies, the colour scale is shown with the colour bar. The orange star marker symbolises the known 42 WHL clusters of galaxies in the identified Saraswati structure. \textbf{Lower Panal:} figure shows the 3D distribution of galaxies and 42 WHL galaxy clusters comprised in the MITRO-identified Saraswati structure. Spheres represent the known WHL-galaxy cluster and the corresponding colour represents the mass of the cluster in log scale as shown with the colour bar. The radius of each sphere is proportional to its radius $r_{200}$ corresponding to the mass $M_{200}$ defined using eq. (2) in \cite{Wen2012}.}
\end{figure*}

Our algorithm spotted a total of 53 structures within the specified analysis region of the sky. The MITRO-identified structures are shown in Fig.~\ref{All_SDSS_structures}, each with different colour and a unique cluster ID (CID). Unique ID has been assigned in ascending order of the total weighted mass of the structures. As expected, the largest and most massive, consequently the first structure (CID-1) in the list is the Saraswati-supercluster. Our algorithm identified this supercluster with an extension of $\sim210$~Mpc (co-moving) located at R.A. $354.41^{\circ}$, and Dec. $0.267^{\circ}$ (LGP centre) and at the mean redshift of $z = 0.277$. We found 384 member galaxies in MITRO identified Saraswati supercluster (Fig.~\ref{Saraswati_cluster3d}) with an estimated average co-moving linking length of $\approx 9.4$ Mpc. The entire structure encloses $\approx 6.47 \times 10^{5}$ Mpc$^{3}$ co-moving volume or $\approx 2.23 \times 10^{5}$ Mpc$^{3} h^{-3}$ co-moving volume in scale of Hubble parameter having an effective mass of $1.86 \times 10^{17}$ M$_{\odot}$ and houses 42 known WHL\footnote{WHL is the short-hand of the initials of the three authors; named Wen Z. L., Han J. L., and Liu F. S. who identify 132,684 galaxy clusters using the photometric redshifts of galaxies from the SDSS in the redshift range of $0.05 < z < 0.8$ using FoF algorithm and constraint on cluster richness (ratio between the total r-band luminosity within the radius $R_{200}$ to the evolved characteristic luminosity of galaxies in the r band). The detailed cluster detection algorithm can be found in \citet{Wen2009, Wen2012}}-galaxy clusters (see Tab.~\ref{tab:table2}). The enclosed co-moving volume is computed with the SciPy Quick-hull Algorithm \cite{Barber1996} for the Convex Hulls as discussed in Sec.~\ref{halo_over_density}.

Previously, \citet{Bagchi2017} (JB17) in their discovery paper reported the Saraswati supercluster to be a structure spanning $\sim200$~Mpc roughly centring at Abell 2631 and at a mean redshift of $z=0.28$. They implemented the FoF algorithm to identify the structure and the linking length of $\ell_0=12$~Mpc was so chosen by them that they could identify the maximum number of FoF clusters in the searched volume with at least 10 galaxies in each. Within the filamentary structure of Saraswati, they found 43 WHL clusters and the computed volume of the Convex Hull of these 43 cluster points is reported to be $V\approx2.05\times10^5$ ~Mpc$^3$. The authors estimated the mass of Saraswati using a very crude method of summing the so-called bound halo mass, $M_{5.6}$ of 43 WHL clusters associated with the structure which they reported as $M_{ss}\sim2\times10^{16}$~M$_{\odot}$. The authors further mentioned that their estimation is strictly a lower limit as a lot of mass may still remain uncounted.

This is noteworthy to mention that our findings are fairly consistent with the JB17, especially the extent, the enclosed co-moving volume in the Hubble scale, the average linking length and the number of identified WHL clusters within the structure. Interestingly, the LGP centre and the average redshift for this supercluster, as identified by MITRO code, are found to be very close to the most massive galaxy cluster inside the structure, i.e., Abell 2631 ($M_{500} \approx 10^{15}$ M$_{\odot}$). However, the estimated total mass and the computed over-density for the structure do not compare well. The difference may be attributed to the method of estimation of mass as well as enclosed volume. We considered the co-moving volume and estimated the weighted mass of the member galaxies that well accounts for the uncounted mass. Whereas, JB17 computed the over-density contrast by considering the sum of the turn-around mass of each WHL-clusters and the Hubble parameter scaled co-moving volume of the enclosed structure. The over-density factor (over critical density) reported by JB17 is $\rho/\rho_{c}= 1.17$, while our estimation is slightly higher $\rho/\rho_{c}= 1.62$. However, both the numbers are far lesser than the required critical value of $\rho/\rho_{c}= 2.36$ that is necessary for the supercluster to remain gravitationally bound. Further trials revealed that, at the place of Saraswati structure, theoretically bound halos can only be found when searched with $\Delta_{SDSS} = 3.78$, however, the structure does not remain intact, rather splits into three parts with this choice of $\Delta_{SDSS}$ (more on this in Appendix~\ref{Apx:Delta_Choice}). This is very important to mention here that the density contrast of $\rho/\rho_{c}= 2.36$ actually comes from the spherical collapse model which may not be fully applicable to our case of a real unstructured supercluster. Moreover, the finding could also be very much dependent on the limitations of the current observations and therefore opens up the possibility of a further detailed study on this matter.

\begin{table*}
\caption{\label{table:clusterlist}List of 53 MITRO identified structures in JB17-Saraswati wedge region  (`LEGACY', `BOSS', and `SOUTHERN' program of SDSS-III DR12). ``*" for indicating the Saraswati supercluster with CID 1.}
\begin{ruledtabular}
\begin{tabular}{c c c c c c c c c c}
CID & Mass & $z$ & R.A. & dec. & Richness & $\frac{\overline{\rho}}{\rho_{c}}$ & $R_{max}$ & LLS & $\bar{l}$\\
    & $(10^{15} M_{\odot})$ &   & (deg.) & (deg.) &   &   & (Mpc) & (Mpc) & (Mpc)\\
\hline
1*& 185.602 & 0.277 & 354.409 & 0.267 & 384 & 1.619 & 124.976 & 226.306 & 9.397 \\ 
2 & 75.259 & 0.3 & 352.942 & 0.069 & 92 & 3.074 & 71.179 & 112.115 & 11.204 \\ 
3 & 59.827 & 0.302 & 1.018 & -0.009 & 53 & 4.111 & 83.19 & 138.652 & 12.276 \\ 
4 & 54.982 & 0.33 & 336.031 & -1.002 & 40 & 7.168 & 96.034 & 96.034 & 13.122 \\ 
5 & 50.929 & 0.318 & 4.709 & -0.094 & 41 & 3.546 & 74.782 & 111.204 & 12.518 \\ 
6 & 49.397 & 0.307 & 6.554 & 0.084 & 48 & 2.756 & 78.104 & 116.96 & 11.992 \\ 
7 & 43.796 & 0.253 & 359.801 & 0.251 & 99 & 1.951 & 77.811 & 116.988 & 9.256 \\ 
8 & 34.449 & 0.235 & 340.97 & -0.123 & 42 & 5.085 & 45.148 & 75.611 & 11.802 \\ 
9 & 30.623 & 0.321 & 351.079 & -0.029 & 20 & 8.501 & 44.268 & 66.634 & 13.792 \\ 
10 & 30.237 & 0.232 & 6.156 & -0.026 & 45 & 3.134 & 53.775 & 95.147 & 10.786 \\
\end{tabular}

\begin{tablenotes}
\small
\item \textit{Note.} Column 1: Structure unique ID. Column 2: Total effective mass of the structure (Stellar mass + surrounding missing matter (baryonic as well as dark matter)). Column 3: Structure redshift. Column 4: RA (J2000) of structure (degree). Column 5: Declination (J2000) of structure (degree). Column 6: Richness-number of galaxies (according to the SDSS-III DR12). Column 7: Structure over-density with respect to the critical density at the structure's redshift. Column 8: Maximum radius of the structure (co-moving). Column 9: co-moving largest linear size of the structure. Column 10: Average linking lengths in structure.
\item (This table is available in its entirety in a machine-readable form.)
\end{tablenotes}

\end{ruledtabular}
\end{table*}

\begin{table*}
\caption{\label{table:candidate_supercluster}List of `candidate-superclusters' of galaxies in JB17-Saraswati wedge region  (`LEGACY', `BOSS', and `SOUTHERN' program of SDSS-III DR12)}
\begin{ruledtabular}
\begin{tabular}{c c c c c c c c c c c c}
CID & Mass & $z$ & R.A. & dec. & Richness & $\frac{\overline{\rho}}{\rho_{c}}$ & $R_{max}$ & LLS & $\bar{l}$ & No. of WHL- & Shape\\
    & $(10^{15} M_{\odot})$ &   & (deg.) & (deg.) &   &   & (Mpc) & (Mpc) & (Mpc)& Clusters\\
\hline
2 & 75.259 & 0.3 & 352.942 & 0.069 & 92 & 3.074 & 71.179 & 112.115 & 11.204 & 6 & circular \\ 
3 & 59.827 & 0.302 & 1.018 & -0.009 & 53 & 4.111 & 83.19 & 138.652 & 12.276 & 5 & elongated \\ 
4 & 54.982 & 0.33 & 336.031 & -1.002 & 40 & 7.168 & 96.034 & 96.034 & 13.122 & 4 & elongated \\ 
5 & 50.929 & 0.318 & 4.709 & -0.094 & 41 & 3.546 & 74.782 & 111.204 & 12.518 & 4 & circular \\ 
6 & 49.397 & 0.307 & 6.554 & 0.084 & 48 & 2.756 & 78.104 & 116.96 & 11.992 & 8  & wall-like \\ 
10 & 30.237 & 0.232 & 6.156 & -0.026 & 45 & 3.134 & 53.775 & 95.147 & 10.786 & 4 & wall-like \\ 
13 & 23.894 & 0.274 & 343.581 & 0.688 & 36 & 4.242 & 62.194 & 106.461 & 10.583 & 3 & elongated \\ 
\end{tabular}

\begin{tablenotes}
\small
\item \textit{Note.} Column 1: Structure unique ID. Column 2: Total effective mass of the structure (Stellar mass + surrounding missing matter (baryonic as well as dark matter)). Column 3: Structure redshift. Column 4: RA (J2000) of structure (degree). Column 5: Declination (J2000) of structure (degree). Column 6: Richness-number of galaxies (according to the SDSS-III DR12). Column 7: Structure over-density with respect to the critical density at the structure's redshift. Column 8: Maximum radius of the structure (co-moving). Column 9: co-moving largest linear size of the structure. Column 10: Average linking lengths in structure. Column 11: Number of known WHL-galaxy clusters in the structure (Tab.~\ref{tab:table2}). Column 12: Morphology/Shape of structure in declination projected/suppressed 2-\textit{dim} plane.
\item elongated: stretched (or LLS) along the line-of-sight (LoS), wall-like: stretched (or LLS) along the angular plane and squeezed along the LoS (along the redshift-axis), and circular: stretched almost equally in both directions.
\end{tablenotes}

\end{ruledtabular}
\end{table*}

In addition to the Saraswati supercluster, we identify another 52 structures by applying the same over-density threshold inside the specified analysis region. The associated physical properties, such as the total effective mass, location, richness, over-density ratio, maximum radius, and largest linear size (LLS) of all the identified structures are listed in Table~\ref{table:clusterlist} (see Tab.~\ref{table:clusterlist_all}). The location mentioned in the table is the LGP centres of each of the structures. It has been found that the final over-density values of 51 out of 52 of these structures are above the threshold value of a bound supercluster. However, many of them are not massive or large enough in extent to call them as superclusters. We put a reasonable constraint on the physical parameters to identify the most probable candidate superclusters in the list. We classified the structures as candidate superclusters if the object has a mass greater than $10^{16}$ M$_{\odot}$, i.e., roughly $\sim$ 10 times $10^{15}$ M$_{\odot}$, the observed massive galaxy clusters and LLS of $\gtrsim 90$ Mpc, the homogeneity scale at which the statistical properties of the Universe have translational invariance \cite{Bharadwaj1999, Ntelis2017}. By applying these constraints, in total, seven MITRO-identified structures were found to qualify as the `candidate-superclusters', and are given in Table~\ref{table:candidate_supercluster}. Based on their morphology on the declination-suppressed 2-D plane, we also defined the shape of these structures which are given in column 12 of Table~\ref{table:candidate_supercluster}. Interestingly, all the candidate-superclusters have an over-density factor of bound structures and encompass at least 3 known WHL-galaxy clusters (respectively,  columns 7 \& 11 of Tab.~\ref{table:candidate_supercluster}). \\

\section{\label{summary}Summary and conclusions}
Large-scale structures in the Universe are known to comprise of groups and clusters of galaxies embedded in a filamentary structure of Dark matter forming the cosmic web. Baryonic matter flows and drains through these DM channels and gets trapped at the large gravitationally bound structures, such as clusters and groups of galaxies that form at the nodes or the crossroads of these filaments. Studies that revealed these structures are majorly based on cosmological large-volume simulations and large-scale galaxy redshift surveys. The most important step in these studies is the appropriate identification of structures in the Universe. Whether the cosmological simulations or the galaxy surveys, the unit element in the data set is a point source, either a DM particle or a galaxy. The known information about them is basically the position and the velocity. While, mass information is readily available in the case of simulations, in galaxy surveys, mass can just be a derived quantity. 

The bound structures in these studies are so far identified using various group finding codes, mostly based on the friend of friends (FoF) or spherical over-density (SO) algorithms. Although, the main goal of these codes is to identify gravitationally bound structures, unfortunately, the mass information (i.e., the origin of the strongest binding force at the large scales), has hardly been effectively used by these codes. Moreover, while both from theory and observations, it is well known that the bound structures can best be formed at some particular mass over-dense condition and real structures are ideally never spherical, the methods used so far either constrain the over-density or the real unstructured geometry. Although, these are very essential parameters to extract accurate structures-mass information and precisely constrain the cosmological models of the Universe, almost no attempt has been made so far to retain these important parameters together while formulating the grouping algorithms. 

In this paper, we present our algorithm called the Measure of Increased Tie with gRavity Order (MITRO) that takes care of all the above-mentioned relevant features as well as ensures the bound structures using the physical quantities, mainly mass and the total energy information. The proposed MITRO algorithm adopts a noble approach of clustering to mitigate a few shortcomings of the basic FoF and SO algorithms strengthening the foundation of clustering or halo-finding methods. Here we briefly summarise the salient features of this proposed algorithm below.

\begin{itemize}
    \item[--] Our major goal was to retain the real geometry, as well as, to identify physically bound cosmological structures in galaxy surveys or simulated cosmological volume. To achieve this, for the first time, we used the dominant gravitational force to determine the linking length between any two elements of a group. In the proposed algorithm, the arm-length has been computed at a user-defined mass over-density (see Sec.~\ref{armlen}) for each element depending on their individual mass. This leads to having a distinct linking length (addition of two arm-lengths) for each unique pair of elements. This is notably different than the usual FoF method, where a single linking length is used for each and every element, usually determined statistically as the scaled mean particle distance in the examined volume and has no dependence on the mass (i.e. gravitational force) of the elements. \\

\item[--] The MITRO-identified halos are the unstructured connected group of spherical elements having a radius defined by their arm-length. The over-density of the MITRO halos thus is determined by the collection of these spherical member elements and eventually constrained by the predefined over-density factor. This is quite different from SO halos, whose over-density is the same as the predefined over-density factor. Moreover, the use of over-density as the parameter to determine the arm length allowed us to constrain or at least put an upper limit to the over-density of the final gravitationally bound structure searched using the MITRO algorithm. It may also resolve the existing issue of the over-density relation  of FoF halos on the mass resolution in simulation setups.\\

\item[--] We demonstrate that the proposed algorithm has been able to successfully identify the unstructured DM halos roughly within the pre-defined density threshold which can not be achieved simultaneously by usual FoF or SO-based methods. This method is therefore a significant improvement as it identifies appropriate over-dense regions without imposing the spherical geometry, moreover, it uses physically relevant linking parameters, as demonstrated through various tests (see Sec.~\ref{over-density_ratio}, ~\ref{compactness_result}) presented in this article. At the same time, the algorithm adopted the important features from both SO and FoF methods such as over-density and percolation.\\ 

\item[--] We could demonstrate that the MITRO-identified halos have over-density contrast at least half of the pre-defined one. This enabled us to define a new parameter called compactness. This parameter is the indicator of how closely the elements of MITRO halos are packed or how stiff the distribution is. Eventually, it reveals the cuspiness of the density profile divulging information about the surrounding environment in which the particular halo is evolving. No similar quantity can be defined in usual SO-halos, as all halos have the same predefined over-density and consequently, compactness is $C_{p} = 1$ for all. On the other hand, since FoF halos are not bound by any over-density contrast linked to the pre-defined linking length, computing compactness is not possible. Our analysis shows compactness of MITRO-identified halos varies as $C_{p} \geq 1$. Here, the higher $C_{p}$ values hint at the lack of matter in the vicinity of evolving halos for the searched $\Delta_{arm}$.\\

\item[--] The critical test that these clustering codes face is while analysing galaxy redshift survey data. Mainly because, such surveys do not directly provide any mass information, as well as, because of missing matter issues due to detection limitations. The usual SO method, which should primarily find the peak of matter density in the search volume, would fail due to the lack of mass information. On the other hand, the application of FoF may result in incorrect grouping because of the use of an unphysical single linking length for all elements in the search volume. In the line of sight, dealing with the redshift distortion effect in the usual way is rather complicated as it requires dedicated cosmological simulations to estimate the equivalent elongation factor. Instead, we overcome the issue considerably using a generalised theoretical approach. Moreover, such an approach helped us to compute the required elongation for all kinds of density contrast, making life easier. 

For better mass estimation of each galaxy, the colour-derived stellar mass has been used in our code. Additionally, the remaining unaccounted matter due to undetected galaxies, IGM and DM content, has been estimated by considering the cosmological mean-matter density at each redshift bin within which it obeys the predicted theoretical mean-matter density. Such an assumption automatically adjust the computed arm length of each element and eventually compensate for the missing mass information. Our theoretical approach helped us to estimate an appropriate elongation factor along the line of sight to overcome the RSD effect. The elongation factor has been determined by assuming the virialization of the halos while computing the peculiar velocities of the galaxies in our setup. Interestingly, the estimated LoS elongation factor of $\sim$ 5-15 (for over-density density $\Delta \sim 180$) in our study is found to be fairly consistent with many earlier studies.\\

\item[--] Finally, we examined the sky region centered at the recently discovered SARASWATI supercluster, by running the MITRO algorithm on SDSS-DR12 data. Our code has successfully identified the supercluster extending $\sim210$~Mpc (co-moving) at a mean redshift of $z = 0.277$ having $384$ member galaxies and housing $42$ WHL galaxy clusters. The findings are largely consistent with the previous study by JB17. However, with a more robust and different mass estimation formalism that better accounts for the missing matter, we found the structure to be of $M_{SCL}=1.86 \times 10^{17}$ M$_{\odot}$. This is an order higher than the previously reported mass and makes it so far the most massive identified supercluster. Furthermore, by setting reasonable criteria, we report seven more MITRO-identified `candidate superclusters' in the field of SARASWATI, indicating that gravitationally bound structures larger than the usual galaxy clusters are not extremely rare. Thus, in the era of increasingly large galaxy redshift surveys (SDSS, Euclid, LSST, etc.), our proposed algorithm MITRO may become a very much useful tool in unravelling the structure formation, constraining the cosmological models, and subsequently understanding the nature of dark energy in the evolutionary history of the Universe.
\end{itemize}

\section*{Acknowledgements}
PG acknowledges the financial support from CSIR, Govt. of India for Senior research fellowship (File No. 09/137/0574/2018 EMR-I). PG wants to thank \textit{Shishir Sankhyayan} and \textit{Shahdab Alam} for helpful discussions on SDSS data. PG also wants to thank \textit{Surhud More} for the useful conversation on the FoF algorithm. SP wants to thank DST for funding his research group through INSPIRE Faculty scheme (IF12/PH-44). The authors want to thank the referee(s) for their useful comments and suggestions. The authors are also thankful to the Inter-University Centre for Astronomy and Astrophysics (IUCAA) for providing the HPC facility. Computations described in this work were performed using the Enzo code developed by the Laboratory for Computational Astrophysics at the University of California in San Diego (http://lca.ucsd.edu).

Funding for SDSS-III has been provided by the Alfred P. Sloan Foundation, the Participating Institutions, the National Science Foundation, and the U.S. Department of Energy Office of Science. The SDSS-III website is \href{http://www.sdss3.org/}{http://www.sdss3.org/}. SDSS-III is managed by the Astrophysical Research Consortium for the Participating Institutions of the SDSS-III Collaboration including the University of Arizona, the Brazilian Participation Group, the Brookhaven National Laboratory, Carnegie Mellon University, the University of Florida, the French Participation Group, the German Participation Group, Harvard University, the Instituto de Astrofisica de Canarias, the Michigan State/Notre Dame/JINA Participation Group, Johns Hopkins University, Lawrence Berkeley National Laboratory, Max Planck Institute for Astrophysics, Max Planck Institute for Extraterrestrial Physics, New Mexico State University, New York University, Ohio State University, Pennsylvania State University, University of Portsmouth, Princeton University, the Spanish Participation Group, University of Tokyo, University of Utah, Vanderbilt University, University of Virginia, University of Washington, and Yale University.

The CosmoSim database used in this paper is a service by the Leibniz Institute for Astrophysics Potsdam (AIP). The authors gratefully acknowledge the Gauss Centre for Supercomputing e.V. (www.gauss-centre.eu) and the Partnership for Advanced Supercomputing in Europe (PRACE, www.prace-ri.eu) for funding the MultiDark simulation project by providing computing time on the GCS Supercomputer SuperMUC at Leibniz Supercomputing Centre (LRZ, www.lrz.de).

Finally, the authors want to acknowledge the developers of the following Python packages used extensively in this work: $scipy$ \cite{Jones2001}, $numpy$ \cite{van2011} and $matplotlib$ \cite{Hunter2007}

\bibliography{prd_v1}



\appendix

\section{MITRO implementation on Simulated mock galaxy sample}\label{MITRO_on_Mock_Sample}
\subsection{Mock Galaxy Sample}

To investigate and compare the halo mass function (HMF) of MITRO and standard FoF-identified galaxy clusters with respect to a theoretical HMF, we utilised a large-volume mock galaxy sample produced by \citet{Cora2018}. This Semi-Analytical Galaxy (\texttt{SAG}) catalogue uses the DM halos information as extracted from a cosmological DM simulation (\texttt{MultiDark} simulation MDPL2), which is a part of the \texttt{CosmoSim} database \cite[see][for more details]{Cora2018}. MDPL2 simulation follows the evolution of $3840^{3}$ particles within a box of $1$ Gpc/h each side, with a flat $\Lambda$CDM cosmology characterised by Planck cosmological parameters: $(\Omega_{\Lambda},\; \Omega_{m},\; \Omega_{b}, \; n_{s})$ = $(0.693,\; 0.307,\; 0.048,\; 0.96)$, $h$ = $H_{0}/(100\; \rm{km}\;\rm{s}^{-1} \rm{Mpc}^{-1})$ = 0.678, and DM particle mass resolution $\sim 1.5 \times 10^{9}$ M$_{\odot}/h$. A few realisations of MDPL2 simulation is publicly available\footnote{\href{https://www.cosmosim.org/metadata/mdpl2/}{https://www.cosmosim.org/metadata/mdpl2/}} on the \texttt{CosmoSim} database\footnote{\href{https://www.cosmosim.org/}{https://www.cosmosim.org/}}.

We prepared a sub-catalogue of Semi-Analytical Galaxies from the (\texttt{SAG}) catalogue\footnote{\href{https://www.cosmosim.org/metadata/mdpl2/sag/}{https://www.cosmosim.org/metadata/mdpl2/sag/}} by extracting data from \texttt{CosmoSim} database within the central volume of ($400$ Mpc/$h$)$^3$ at redshift $z = 0.0$. The following information was recorded in a data file: halo-mass, x-position, y-position, z-position, the absolute magnitude of galaxies with respect to SDSS Camera $r$-band response function (assuming airmass = 1.3, in rest frame), mass of stars in the spheroid/bulge, mass of stars in the disk, mass of gas in the spheroid/bulge, mass of gas in the disk, hot gas mass, mass of central black hole. Here, the total mass of each galaxy is the sum of all extracted masses. Finally, we put a constraint on the absolute magnitude of galaxies in order to mimic the observations, i.e., $r$-band absolute magnitude $<-17$ (as obtained in our SDSS sample). Such an exercise of removing the galaxies on the basis of  $r$-band absolute magnitude, from the examined volume, introducing the missing matter problem in this mock galaxy sample. In the final list, a total of 4243667 galaxies were retained in the selected volume.

\subsection{MITRO galaxy clusters identified from the SAG catalogue}
\begin{figure}
   \includegraphics[width=\columnwidth]{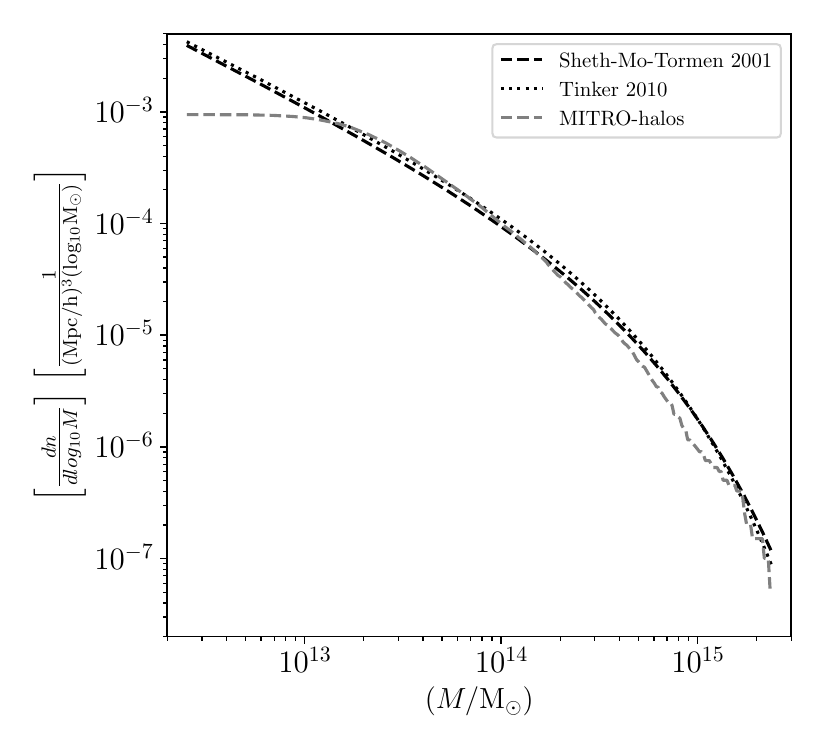}
   \caption{\label{fig:Mitro_HMF_SAG} Halo mass function of MITRO-galaxy clusters identified from SAG catalogue.}
\end{figure}

We implemented our proposed MITRO algorithm following the steps mentioned in Sec.~\ref{missingmatter} for accounting for the missing matter within the selected volume and accordingly, the weight factor ($w$) was computed. Finally, the over-dense regions, in this case, the galaxy clusters, are identified with the MITRO algorithm, assuming arm-length corresponding to $\Delta_{arm} = 200$ and weighted mass of each galaxy as $w \times M^{gal}_{i}$. In the given sample with the above-stated parameters and the Eq.~\ref{eq:eq5}, the $\{max.,\; min.,\; mean,\; median\}$ arm-length would be $\{2645.27,\; 25.14,\; 122.14,\; 101.41\}$ kpc. Since the given sample is produced from cosmological simulations is not affected by the redshift distortion and thus the elongation factor need not be considered. The search for the over-dense regions is therefore done in spherical kernels around the galaxies instead of using a cylindrical kernel. Furthermore, with the constraints on minimum DM particles, $N_{min} = 10$, and on the total mass of galaxy cluster $M_{halo} \geq 10^{12}$ M$_{\odot}$, MITRO-algorithm identified 18821 galaxy clusters with the massive most having mass $2.4 \times 10^{15}$ M$_{\odot}$. The corresponding HMF has been shown in Fig.~\ref{fig:Mitro_HMF_SAG}, which is in good agreement with the standard Sheth-Tormen 2001 and Tinker 2010 HMFs $\geq 10^{13}$ M$_{\odot}$.

\subsection{FoF galaxy clusters identified from SAG catalogue}

\begin{figure}
   \includegraphics[width=\columnwidth]{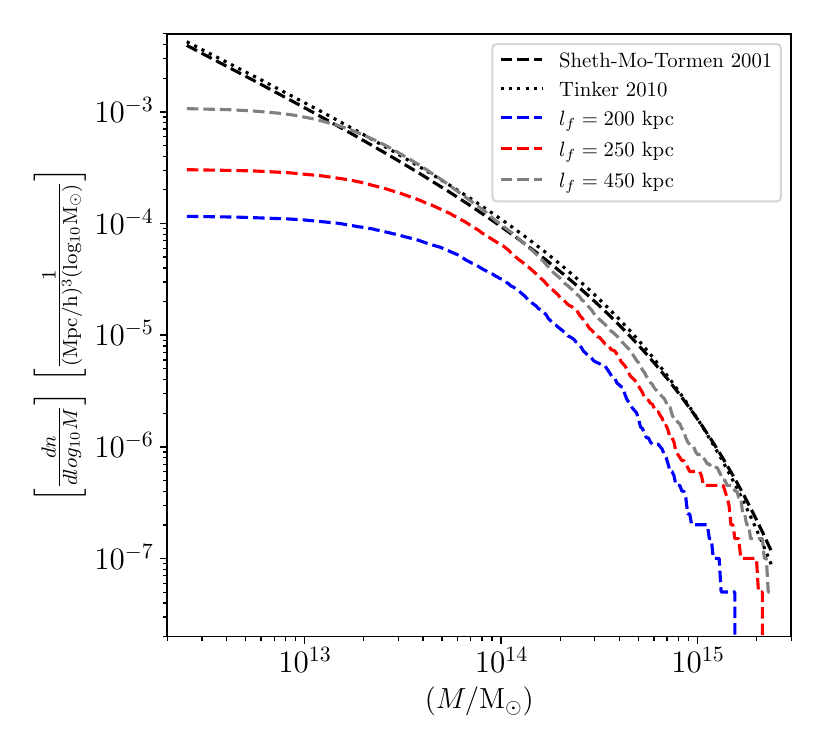}
   \caption{\label{fig:FoF_HMF_SAG} Halo mass function of FoF galaxy clusters identified from SAG catalogue. Different coloured dashed lines correspond to different linking lengths as mentioned in the legend.}
\end{figure}

Since there is no direct (co-)relation between the linking length and searched over-density threshold, the linking length ($l_{f}$) in the FoF algorithm is still a quest. Thus, for a meaningful comparison between FoF and MITRO, we chose two statistically important pointers, i.e., mean and median of the arm-lengths of all objects in the MITRO algorithm when searched at $\Delta\_{arm} = 200$. Since with this approximation, the final linking length should be equivalent to twice the chosen arm length, the equivalent linking lengths would be $l_{f_{mean}} \sim 250$~kpc and $l_{f_{median}} \sim 200$~kpc, respectively. The produced FoF halo-mass function (Fig.~\ref{fig:FoF_HMF_SAG}) shows they do not match well with the expected halo-mass function of Sheth-Tormen 2001 nor Tinker et al. 2010. The number of clusters identified by FoF was also found to be considerably less (2347 and 6119, respectively) in comparison to MITRO (i.e., 18821). This lead us to do a further iterative test that resulted in a linking length of $l_{f} = 450$~kpc, where FoF could roughly able to produce a match with the halo-mass function of Tinker 2010 and Sheth-Tormen 2001. This certainly indicates the real strength of the MITRO algorithm as MITRO produces desired results without the need for an iterative trial. In addition, our algorithm also provides a tentative over-density threshold for the identified structure i.e., $\gtrsim \Delta_{arm}/2 \sim 100$.\\

\section{Structures in SDSS selected wedge near Saraswati region}
\label{Apx:SDSS}
\subsection{Choice of over-density search in SDSS sample}\label{Apx:Delta_Choice}
\begin{figure*}
   \includegraphics[width=\columnwidth]{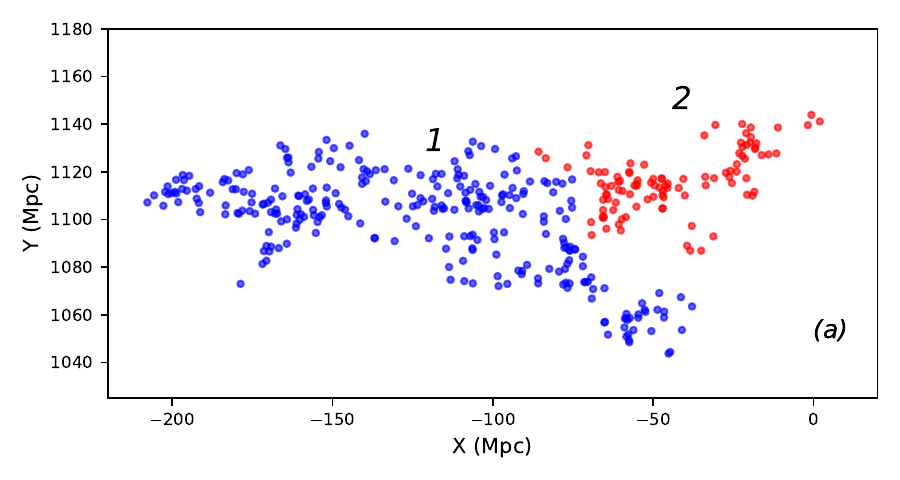}
   \includegraphics[width=\columnwidth]{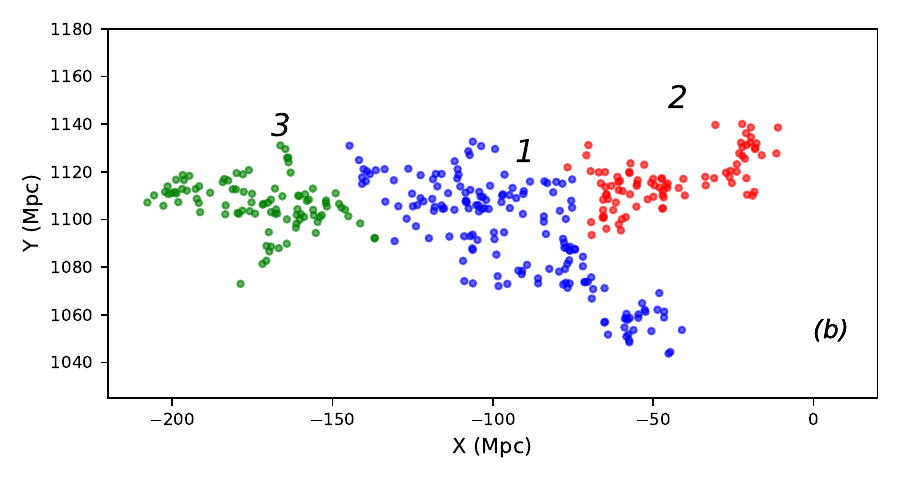}
   \caption{{\bf Panel~a}: MITRO identified structure at the position of Saraswati supercluster splits into two parts with the choice of $\Delta_{SDSS} = 2.60$. Part~2 (red) parts away from the main structure. {\bf Panel~b:} The same structure when searched at $\Delta_{SDSS} = 3.78$, splits into three distinct parts.}\label{fig:Delta_SDSS_10-60perct}
\end{figure*}

We carried out an iterative test (each step of 10\% increment) to understand the sensitivity of the $\Delta_{SDSS}$ parameter. We found, even if we increase the value just by 10\% (i.e.,  $\Delta_{SDSS}$ = 2.60) of the chosen value (i.e., $\Delta_{SDSS}$ = 2.36), the Saraswati supercluster splits into two parts. The wing-like structure on the right side of the supercluster gets detached (see Fig.~\ref{fig:Delta_SDSS_10-60perct}a). Likewise, when we increase the value by 60\% of 2.36 ($\Delta_{SDSS}$ ~ 3.78), the structure splits into three parts (left wing, centre part + lower tail, right-wing, see Fig.~\ref{fig:Delta_SDSS_10-60perct}b).
However, similar to at $\Delta_{SDSS} = 2.36$, the density threshold of identified two splitted parts at $\Delta_{SDSS} = 2.60$ are also not above the theoretical limit of gravitationally bound structures. It is only at $\Delta_{SDSS} = 3.78$ that all three splitted structures become gravitationally bound.

\vspace{-10pt}

\begin{table*}
\caption{\label{table:clusterlist_all}List of 53 MITRO identified structures from JB17-Saraswati-region (`LEGACY', `BOSS', and `SOUTHERN' program of SDSS-III DR12)}
\begin{ruledtabular}
\begin{tabular}{c c c c c c c c c c}
CID & Mass & $z$ & R.A. & dec. & Richness & $\frac{\overline{\rho}}{\rho_{c}}$ & $R_{max}$ & LLS & $\bar{l}$\\
    & $(10^{15} M_{\odot})$ &   & (deg.) & (deg.) &   &   & (Mpc) & (Mpc) & (Mpc)\\
\hline
1*& 185.602 & 0.277 & 354.409 & 0.267 & 384 & 1.619 & 124.976 & 226.306 & 9.397 \\ 
2 & 75.259 & 0.3 & 352.942 & 0.069 & 92 & 3.074 & 71.179 & 112.115 & 11.204 \\ 
3 & 59.827 & 0.302 & 1.018 & -0.009 & 53 & 4.111 & 83.19 & 138.652 & 12.276 \\ 
4 & 54.982 & 0.33 & 336.031 & -1.002 & 40 & 7.168 & 96.034 & 96.034 & 13.122 \\ 
5 & 50.929 & 0.318 & 4.709 & -0.094 & 41 & 3.546 & 74.782 & 111.204 & 12.518 \\ 
6 & 49.397 & 0.307 & 6.554 & 0.084 & 48 & 2.756 & 78.104 & 116.96 & 11.992 \\ 
7 & 43.796 & 0.253 & 359.801 & 0.251 & 99 & 1.951 & 77.811 & 116.988 & 9.256 \\ 
8 & 34.449 & 0.235 & 340.97 & -0.123 & 42 & 5.085 & 45.148 & 75.611 & 11.802 \\ 
9 & 30.623 & 0.321 & 351.079 & -0.029 & 20 & 8.501 & 44.268 & 66.634 & 13.792 \\ 
10 & 30.237 & 0.232 & 6.156 & -0.026 & 45 & 3.134 & 53.775 & 95.147 & 10.786 \\ 
11 & 28.789 & 0.293 & 5.959 & 0.568 & 43 & 4.131 & 51.605 & 75.443 & 10.554 \\ 
12 & 24.32 & 0.317 & 339.708 & 1.078 & 19 & 10.283 & 52.201 & 56.732 & 12.956 \\ 
13 & 23.894 & 0.274 & 343.581 & 0.688 & 36 & 4.242 & 62.194 & 106.461 & 10.583 \\ 
14 & 23.052 & 0.312 & 347.422 & 1.104 & 18 & 10.979 & 50.394 & 57.309 & 13.238 \\ 
15 & 22.924 & 0.292 & 347.077 & -0.187 & 32 & 3.93 & 47.131 & 78.523 & 10.811 \\ 
16 & 22.121 & 0.259 & 5.826 & 0.192 & 46 & 3.777 & 36.005 & 65.306 & 9.781 \\ 
17 & 20.143 & 0.31 & 345.581 & 0.603 & 17 & 13.993 & 54.98 & 71.364 & 12.427 \\ 
18 & 19.328 & 0.315 & 337.573 & 0.68 & 18 & 10.365 & 35.539 & 48.453 & 12.229 \\ 
19 & 18.484 & 0.268 & 345.786 & -0.92 & 26 & 7.053 & 39.845 & 50.391 & 10.955 \\ 
20 & 17.365 & 0.294 & 336.605 & 0.893 & 21 & 11.901 & 24.156 & 42.673 & 11.278 \\ 
21 & 16.918 & 0.259 & 9.098 & 0.533 & 30 & 4.835 & 48.112 & 69.719 & 10.01 \\ 
22 & 16.135 & 0.29 & 344.517 & 0.357 & 20 & 8.2 & 51.161 & 66.819 & 11.507 \\ 
23 & 15.632 & 0.323 & 342.799 & 0.704 & 13 & 7.809 & 45.396 & 53.84 & 12.134 \\ 
24 & 15.374 & 0.3 & 338.308 & 0.131 & 18 & 12.646 & 38.697 & 57.745 & 11.413 \\ 
25 & 15.336 & 0.287 & 350.167 & -1.014 & 20 & 7.672 & 42.938 & 56.825 & 11.433 \\ 
26 & 14.433 & 0.302 & 11.856 & 0.679 & 17 & 7.564 & 32.612 & 49.729 & 11.026 \\ 
27 & 12.616 & 0.298 & 338.9 & -1.077 & 11 & 14.114 & 47.898 & 47.898 & 12.767 \\ 
28 & 12.294 & 0.313 & 10.766 & 0.049 & 10 & 8.875 & 33.85 & 49.348 & 12.517 \\ 
29 & 12.009 & 0.326 & 358.504 & -0.58 & 12 & 11.029 & 26.54 & 46.212 & 11.748 \\ 
30 & 11.338 & 0.259 & 15.307 & -0.719 & 18 & 11.907 & 35.758 & 47.51 & 11.163 \\ 
31 & 10.43 & 0.243 & 15.626 & -0.848 & 15 & 21.887 & 30.041 & 54.335 & 10.623 \\ 
32 & 10.419 & 0.285 & 14.34 & 0.849 & 18 & 16.294 & 27.533 & 40.096 & 10.094 \\ 
33 & 10.063 & 0.256 & 354.491 & 0.776 & 22 & 9.457 & 41.29 & 51.335 & 9.616 \\ 
34 & 9.454 & 0.301 & 348.575 & 0.191 & 13 & 4.96 & 39.153 & 54.615 & 11.118 \\ 
35 & 8.573 & 0.235 & 7.463 & 0.754 & 14 & 10.082 & 42.411 & 55.644 & 10.124 \\ 
36 & 7.844 & 0.242 & 337.205 & -0.95 & 13 & 19.201 & 38.987 & 51.832 & 10.01 \\ 
37 & 7.786 & 0.275 & 339.645 & 0.743 & 11 & 15.349 & 21.771 & 35.626 & 10.918 \\ 
38 & 7.099 & 0.247 & 343.733 & -1.042 & 10 & 22.859 & 21.615 & 27.324 & 11.731 \\ 
39 & 6.968 & 0.254 & 337.905 & 0.259 & 11 & 17.649 & 23.835 & 40.85 & 10.882 \\ 
40 & 6.926 & 0.283 & 341.549 & 0.494 & 11 & 10.472 & 25.987 & 39.252 & 9.881 \\ 
41 & 6.734 & 0.264 & 353.209 & -1.041 & 10 & 15.8 & 24.33 & 34.619 & 10.722 \\ 
42 & 6.578 & 0.272 & 347.447 & 0.055 & 10 & 16.784 & 39.977 & 51.357 & 10.245 \\ 
43 & 6.554 & 0.26 & 344.632 & -0.111 & 13 & 6.266 & 29.5 & 47.385 & 9.828 \\ 
44 & 6.223 & 0.268 & 10.69 & 0.473 & 14 & 6.655 & 27.749 & 47.466 & 9.581 \\ 
45 & 6.075 & 0.251 & 351.795 & -1.139 & 11 & 14.511 & 30.705 & 32.196 & 10.283 \\ 
46 & 5.217 & 0.262 & 13.614 & 0.461 & 10 & 22.064 & 21.594 & 24.636 & 9.712 \\ 
47 & 4.789 & 0.276 & 9.9 & -0.494 & 12 & 10.204 & 24.451 & 34.629 & 8.939 \\ 
48 & 4.751 & 0.252 & 339.093 & -0.514 & 10 & 24.829 & 21.683 & 33.723 & 9.555 \\ 
49 & 4.278 & 0.254 & 339.177 & -0.275 & 12 & 7.481 & 24.232 & 37.156 & 8.338 \\ 
50 & 4.274 & 0.268 & 12.31 & -0.158 & 10 & 7.884 & 22.534 & 37.062 & 8.97 \\ 
51 & 4.271 & 0.252 & 353.213 & 1.174 & 10 & 7.958 & 35.918 & 39.626 & 9.821 \\ 
52 & 3.823 & 0.248 & 3.508 & 0.794 & 12 & 15.294 & 34.126 & 41.634 & 8.599 \\ 
53 & 3.239 & 0.232 & 358.556 & -0.321 & 11 & 7.416 & 25.007 & 36.919 & 8.273 \\ 
\end{tabular}
\end{ruledtabular}
\end{table*}

\squeezetable
\begin{table*}
\caption{\label{tab:table2}WHL clusters in MITRO identified galaxy clusters and superclusters}
\begin{ruledtabular}
\begin{tabular}{c c c c c| c c c c c}
CID & WHL clusters & $z$ & R.A. (deg.) & dec. (deg.) & CID & WHL clusters & $z$ & R.A. (deg.) & dec. (deg.)\\
\hline
1*& WHL J232229.7-002223 & 0.2728 & 350.62363 & -0.37294 &   & WHL J002648.8-001610 & 0.2311 & 6.70325 & -0.26937 \\
  & WHL J232406.1+002147 & 0.278 & 351.02548 & +0.36311 &   & WHL J003130.2-010303 & 0.232 & 7.87587 & -1.05083 \\
  & WHL J232517.4+001639 & 0.2767 & 351.32236 & +0.27741 &   &   &   &   &   \\
  & WHL J232643.3+010803 & 0.2774 & 351.68036 & +1.13423 & 11 & WHL J002400.3+003007 & 0.2923 & 6.00129 & +0.50202 \\
  & WHL J232731.9+005634 & 0.2788 & 351.88278 & +0.94281 &   & WHL J002813.8-002154 & 0.2912 & 7.05762 & -0.36497 \\
  & WHL J232809.2+001109 & 0.2768 & 352.03851 & +0.18593 &   &   &   &   &   \\
  & WHL J232842.6+004939 & 0.2784 & 352.17755 & +0.82753 & 12 & WHL J223818.0+005559 & 0.3173 & 339.57483 & +0.9331 \\
  & WHL J233040.2+003634 & 0.2757 & 352.66733 & +0.6094 &   &   &   &   &   \\
  & WHL J233116.7+004336 & 0.2812 & 352.81964 & +0.72678 & 13 & WHL J224807.9+002040 & 0.2657 & 342.03287 & +0.34439 \\
  & WHL J233126.4+003655 & 0.2739 & 352.8602 & +0.61525 &   & WHL J225424.4+004101 & 0.2738 & 343.60159 & +0.68368 \\
  & WHL J233430.3-005906 & 0.2772 & 353.62631 & -0.98501 &   & WHL J225729.0+003102 & 0.273 & 344.37103 & +0.51717 \\
  & WHL J233530.9-003152 & 0.2764 & 353.87866 & -0.53116 &   &   &   &   &   \\
  & WHL J233558.5-002904 & 0.27 & 353.99393 & -0.48434 & 16 & WHL J002314.8+001158 & 0.2597 & 5.81154 & +0.19944 \\
  & WHL J233739.7+001617 & 0.2772 & 354.41553 & +0.27137 &   & WHL J002358.9-003039 & 0.2575 & 5.99551 & -0.51081 \\
  & WHL J233816.0+000700 & 0.2711 & 354.56683 & +0.1168 &   & WHL J002427.4+002450 & 0.2598 & 6.11418 & +0.41389 \\
  & WHL J233955.0-002558 & 0.2769 & 354.97913 & -0.43282 &   &   &   &   &   \\
  & WHL J234024.5-000535 & 0.2761 & 355.10202 & -0.09299 & 17 & WHL J230408.4-010642 & 0.3147 & 346.03494 & -1.11163 \\
  & WHL J234106.9+001833 & 0.2768 & 355.27872 & +0.30925 &   &   &   &   &   \\
  & WHL J234144.1-004031 & 0.268 & 355.43375 & -0.67539 & 19 & WHL J230227.7-002516 & 0.2647 & 345.61533 & -0.42125 \\
  & WHL J234233.1-001719 & 0.277 & 355.63776 & -0.28874 &   & WHL J230316.7-005114 & 0.2685 & 345.81976 & -0.85396 \\
  & WHL J234335.7+001951 & 0.2694 & 355.89862 & +0.33093 &   & WHL J230348.8-004024 & 0.2662 & 345.95328 & -0.67331 \\
  & WHL J234403.1+001335 & 0.2714 & 356.01273 & +0.22644 &   &   &   &   &   \\
  & WHL J234446.7-000520 & 0.2674 & 356.19455 & -0.0888 & 20 & WHL J222426.4+002737 & 0.2938 & 336.1102 & +0.46019 \\
  & WHL J234519.2-000312 & 0.2665 & 356.3298 & -0.05331 &   & WHL J222542.8+010339 & 0.2952 & 336.42816 & +1.06081 \\
  & WHL J234548.2-010740 & 0.2797 & 356.45068 & -1.12771 &   & WHL J222627.3+005329 & 0.294 & 336.61365 & +0.89143 \\
  & WHL J234604.7-001109 & 0.2665 & 356.5195 & -0.18573 &   &   &   &   &   \\
  & WHL J234623.9+004458 & 0.2746 & 356.59955 & +0.74942 & 21 & WHL J003200.8+001353 & 0.2532 & 8.00329 & +0.23131 \\
  & WHL J234643.2+005003 & 0.2637 & 356.67987 & +0.83429 &   & WHL J003511.8+004348 & 0.262 & 8.79937 & +0.73002 \\
  & WHL J234703.0-000508 & 0.2655 & 356.76236 & -0.08544 &   & WHL J003554.5+000924 & 0.2597 & 8.97729 & +0.15671 \\
  & WHL J234727.0+003140 & 0.2729 & 356.86264 & +0.52773 &   & WHL J003555.1+002123 & 0.2592 & 8.97972 & +0.35646 \\
  & WHL J234727.6-000914 & 0.2639 & 356.86499 & -0.15381 &   & WHL J003614.8+000940 & 0.2569 & 9.06173 & +0.16113 \\
  & WHL J234856.4-005328 & 0.2793 & 357.2352 & -0.8912 &   & WHL J003619.8+002418 & 0.2576 & 9.08261 & +0.40494 \\
  & WHL J234946.6+003846 & 0.2705 & 357.44406 & +0.64598 &   & WHL J003629.8+003520 & 0.2607 & 9.1241 & +0.58899 \\
  & WHL J235016.5+005301 & 0.2767 & 357.56891 & +0.88373 &   & WHL J003711.8+000546 & 0.2588 & 9.29924 & +0.09603 \\
  & WHL J235121.3+003701 & 0.2775 & 357.83884 & +0.61691 &   &   &   &   &   \\
  & WHL J235143.9+003339 & 0.2703 & 357.93292 & +0.56078 & 22 & WHL J225211.0+003531 & 0.2877 & 343.04565 & +0.59185 \\
  & WHL J235229.5+003622 & 0.2698 & 358.12292 & +0.60623 &   &   &   &   &   \\
  & WHL J235301.3-001351 & 0.2779 & 358.25555 & -0.23089 & 24 & WHL J223304.8+000818 & 0.2989 & 338.2699 & +0.13823 \\
  & WHL J235347.5+004859 & 0.2787 & 358.448 & +0.81641 &   &   &   &   &   \\
  & WHL J235424.8+004838 & 0.2805 & 358.60349 & +0.81066 & 25 & WHL J231948.3-010932 & 0.2909 & 349.95108 & -1.15889 \\
  & WHL J235508.9+004512 & 0.2789 & 358.7869 & +0.75341 &   &   &   &   &   \\
  & WHL J235543.8-000014 & 0.2752 & 358.93271 & -0.00393 & 28 & WHL J004051.3+000858 & 0.3132 & 10.2137 & +0.14932 \\
  &   &   &   &    &   &   &   &   &   \\
2 & WHL J232420.2-003229 & 0.293 & 351.08423 & -0.54147 & 29 & WHL J235424.3-001451 & 0.3277 & 358.60135 & -0.24737 \\
  & WHL J232902.7-003847 & 0.2988 & 352.26129 & -0.6463 &   &   &   &   &   \\
  & WHL J232923.7-004856 & 0.3007 & 352.34872 & -0.81546 & 30 & WHL J010154.7-005216 & 0.2616 & 15.47781 & -0.87118 \\
  & WHL J233317.2+000741 & 0.3025 & 353.32187 & +0.12797 &   & WHL J010214.5-010851 & 0.2611 & 15.56023 & -1.14737 \\
  & WHL J233531.2-005320 & 0.2974 & 353.87994 & -0.88882 &   & WHL J010331.1-005932 & 0.2565 & 15.87962 & -0.99216 \\
  &   &   &   &   &   & WHL J010339.2-004508 & 0.2646 & 15.91314 & -0.75235 \\
3 & WHL J000201.8+002049 & 0.3007 & 0.50762 & +0.34701 &   &   &   &   &   \\
  & WHL J000224.7-003253 & 0.2895 & 0.60304 & -0.54798 & 31 & WHL J010229.8-004848 & 0.2375 & 15.62412 & -0.81343 \\
  & WHL J000242.3-001320 & 0.2989 & 0.67632 & -0.22217 &   &   &   &   &   \\
  & WHL J000455.1+003415 & 0.3197 & 1.22949 & +0.57073 & 32 & WHL J005338.1+005515 & 0.2851 & 13.40886 & +0.92088 \\
  & WHL J000825.6+004152 & 0.3124 & 2.10683 & +0.69786  &   & WHL J005434.9+004819 & 0.2847 & 13.6454 & +0.80528 \\
  &   &   &   &   &   & WHL J005526.0+004731 & 0.2846 & 13.85833 & +0.79205 \\
4 & WHL J222456.1-002115 & 0.3154 & 336.23373 & -0.35415 &   &   &   &   &   \\
  & WHL J222523.4-005410 & 0.3204 & 336.3476 & -0.90289 & 33 & WHL J234145.8+010728 & 0.2507 & 355.44077 & +1.12444 \\
  & WHL J222731.7-003945 & 0.3226 & 336.88223 & -0.66258 &   &   &   &   &   \\
  & WHL J222810.8-004848 & 0.327 & 337.04517 & -0.8134 & 34 & WHL J231308.1-010632 & 0.3007 & 348.28375 & -1.10895 \\
  &   &   &   &   &   &   &   &   &   \\
5 & WHL J001848.6-000430 & 0.3184 & 4.70257 & -0.07488 & 35 & WHL J003407.0-001147 & 0.2365 & 8.52936 & -0.19632 \\
  & WHL J002020.7-001925 & 0.3251 & 5.08638 & -0.32355 &   &   &   &   &   \\
  & WHL J002025.3-001216 & 0.3267 & 5.10561 & -0.20444 & 36 & WHL J222652.2-005607 & 0.2483 & 336.71762 & -0.93536 \\
  & WHL J002105.0+001406 & 0.3291 & 5.27082 & +0.23505 &   & WHL J222720.8-005610 & 0.2468 & 336.83655 & -0.93623 \\
  &   &   &   &   &   &   &   &   &   \\
6 & WHL J002200.7-001839 & 0.304 & 5.50288 & -0.31075 & 39 & WHL J223126.2+001743 & 0.2547 & 337.85898 & +0.29532 \\
  & WHL J002334.8+001907 & 0.3084 & 5.89512 & +0.31861 &   &   &   &   &   \\
  & WHL J002557.8+000734 & 0.3076 & 6.49081 & +0.12619 & 40 & WHL J224641.8+002837 & 0.2846 & 341.67407 & +0.47684 \\
  & WHL J002920.4-004032 & 0.3085 & 7.3349 & -0.67557 &   &   &   &   &   \\
  & WHL J003401.5-003314 & 0.3076 & 8.50633 & -0.55392 & 41 & WHL J233608.8-011100 & 0.265 & 354.03653 & -1.18347 \\
  & WHL J003508.9-010803 & 0.3055 & 8.78714 & -1.13429 &   &   &   &   &   \\
  & WHL J003542.0+000158 & 0.3102 & 8.92503 & +0.03278 & 43 & WHL J225830.5-000608 & 0.2587 & 344.62704 & -0.1021 \\
  & WHL J003744.4-011424 & 0.3106 & 9.43509 & -1.23996 &   &   &   &   &   \\
  &   &   &   &   & 44 & WHL J004249.9+001255 & 0.2691 & 10.70773 & +0.21535 \\
7 & WHL J235751.3+002133 & 0.2596 & 359.46381 & +0.3593 &   & WHL J004252.7+004306 & 0.2699 & 10.71962 & +0.71844 \\
  & WHL J235914.3+001405 & 0.2533 & 359.8096 & +0.23484 &   &   &   &   &   \\
  & WHL J235952.8+004155 & 0.2674 & 359.96988 & +0.69856 & 45 & WHL J232708.4-011250 & 0.2517 & 351.7851 & -1.21393 \\
  & WHL J000024.3+003804 & 0.2522 & 0.1013 & +0.63434  &   &   &   &   &   \\
  & WHL J000050.5+004705 & 0.2634 & 0.21052 & +0.78478 & 48 & WHL J223556.1-001411 & 0.2488 & 338.98395 & -0.2364 \\
  & WHL J000117.8-005612 & 0.2473 & 0.32417 & -0.93668 &   & WHL J223622.8-003049 & 0.2488 & 339.09488 & -0.51352 \\
  & WHL J000126.3-000143 & 0.2479 & 0.35969 & -0.02867 &   &   &   &   &   \\
  &   &   &   &   & 49 & WHL J223629.1+000118 & 0.2539 & 339.12143 & +0.02159 \\
8 & WHL J224249.1-001648 & 0.2363 & 340.70477 & -0.2801 &   &   &   &   &   \\
  &   &   &   &   &   &   &   &   &   \\
10 & WHL J001637.4-002739 & 0.2325 & 4.15592 & -0.46084 & 52 & WHL J001229.8+003911 & 0.2505 & 3.12436 & +0.65318 \\
  & WHL J002435.2-001307 & 0.2308 & 6.14679 & -0.21867 &   & WHL J001357.8+004706 & 0.2465 & 3.49083 & +0.78508 \\

\end{tabular}
\end{ruledtabular}
\end{table*}



\end{document}